\magnification=1200
\hsize=15truecm
\vsize=23truecm
\baselineskip 18 truept
\voffset=-0.5truecm
\parindent=0cm
\overfullrule=0pt

\catcode`@=11
%
%---------------------------- \lsim \gsim ----------------------------------
%
%    Simboli di minore o circa uguale, maggiore o circa uguale.
%
\def\lsim{\mathchoice
  {\mathrel{\lower.8ex\hbox{$\displaystyle\buildrel<\over\sim$}}}
  {\mathrel{\lower.8ex\hbox{$\textstyle\buildrel<\over\sim$}}}
  {\mathrel{\lower.8ex\hbox{$\scriptstyle\buildrel<\over\sim$}}}
  {\mathrel{\lower.8ex\hbox{$\scriptscriptstyle\buildrel<\over\sim$}}} }
\def\gsim{\mathchoice
  {\mathrel{\lower.8ex\hbox{$\displaystyle\buildrel>\over\sim$}}}
  {\mathrel{\lower.8ex\hbox{$\textstyle\buildrel>\over\sim$}}}
  {\mathrel{\lower.8ex\hbox{$\scriptstyle\buildrel>\over\sim$}}}
 {\mathrel{\lower.8ex\hbox{$\scriptscriptstyle\buildrel>\over\sim$}}} }

\def\gsu{\raise4pt\hbox{\kern5pt\hbox{$\sim$}}\lower1pt\hbox{\kern-8pt
\hbox{$>$}}~}
\def\lsu{\raise4pt\hbox{\kern5pt\hbox{$\sim$}}\lower1pt\hbox{\kern-8pt
\hbox{$<$}}~}

\def\croce{\displaystyle / \kern-0.2truecm\hbox{$\backslash$}}
\def\lqua{\lower4pt\hbox{\kern5pt\hbox{$\sim$}}\raise1pt
\hbox{\kern-8pt\hbox{$<$}}~}
\def\gqua{\lower4pt\hbox{\kern5pt\hbox{$\sim$}}\raise1pt
\hbox{\kern-8pt\hbox{$>$}}~}
\def\mma{\lower1pt\hbox{\kern5pt\hbox{$\scriptstyle <$}}\raise2pt
\hbox{\kern-7pt\hbox{$\scriptstyle >$}}~}
\def\mmb{\lower1pt\hbox{\kern5pt\hbox{$\scriptstyle >$}}\raise2pt
\hbox{\kern-7pt\hbox{$\scriptstyle <$}}~}
\def\mmc{\lower4pt\hbox{\kern5pt\hbox{$<$}}\raise1pt
\hbox{\kern-8pt\hbox{$>$}}~}
\def\mmd{\lower4pt\hbox{\kern5pt\hbox{$>$}}\raise1pt
\hbox{\kern-8pt\hbox{$<$}}~}
\def\croce{\displaystyle / \kern-0.2truecm\hbox{$\backslash$}}
%
%
%---------------------------  \quadratello ---------------------------------
%
%
\def\quad@rato#1#2{{\vcenter{\vbox{
        \hrule height#2pt
        \hbox{\vrule width#2pt height#1pt \kern#1pt \vrule width#2pt}
        \hrule height#2pt} }}}
\def\quadratello{\mathchoice
\quad@rato5{.5}\quad@rato5{.5}\quad@rato{3.5}{.35}\quad@rato{2.5}{.25} }
%
%------------------------ caratteri grassetto speciali -----------
%
\font\s@=cmss10\font\s@b=cmbx8
\def\reali{{\hbox{\s@ l\kern-.5mm R}}}
\def\m{{\hbox{\s@ l\kern-.5mm M}}}
\def\k{{\hbox{\s@ l\kern-.5mm K}}}
\def\naturali{{\hbox{\s@ l\kern-.5mm N}}}
\def\interi{{\mathchoice
 {\hbox{\s@ Z\kern-1.5mm Z}}
 {\hbox{\s@ Z\kern-1.5mm Z}}
 {\hbox{{\s@b Z\kern-1.2mm Z}}}
 {\hbox{{\s@b Z\kern-1.2mm Z}}}  }}
\def\complessi{{\hbox{\s@ C\kern-1.7mm\raise.6mm\hbox{\s@b l}\kern.8mm}}}
\def\toro{{\hbox{\s@ T\kern-1.9mm T}}}
\def\unity{{\hbox{\s@ 1\kern-.8mm l}}}
%
%------------------------- bold math. it. --------------------------
%
\font\bold@mit=cmmib10
\def\setbmit{\textfont1=\bold@mit}
\def\bmit#1{\hbox{\textfont1=\bold@mit$#1$}}
%
%-----------------------------------------------------------------
\catcode`@=12

\null

\hfill{DFPD/96/TH/26}

\hfill{hep-th/9605115}

\centerline
{\bf DIMENSIONAL REDUCTION OF $U(1)\times SU(2)$ CHERN--SIMONS}
\centerline
{\bf BOSONIZATION: APPLICATION TO THE $t-J$ MODEL}

\vskip 0.5truecm

\centerline{P.A. Marchetti $^*$}
\smallskip
\centerline
{\it Dipartimento di Fisica ``G. Galilei"}
\smallskip
\centerline
{\it Istituto Nazionale di Fisica Nucleare, I--35131 Padova, Italy}

\vskip 0.5truecm

\centerline{Zhao--Bin Su}
\smallskip
\centerline
{\it Institute of Theoretical Physics,}
\smallskip
\centerline
{\it Chinese Academy of Sciences, Beijing 100080, China}

\vskip 0.5truecm

\centerline{Lu Yu}
\smallskip
\centerline
{\it International Centre for Theoretical Physics, I-34100 Trieste, Italy}
\smallskip
\centerline
{\it and Institute of Theoretical Physics,} 
\centerline
{\it Chinese Academy of Sciences,
Beijing 100080, China}

\vskip 0.5truecm

\centerline{\bf ABSTRACT}

\vskip 0.5truecm

We perform a dimensional reduction of the $U(1)\times SU(2)$ Chern--Simons
bosonization and  apply it to the $t-J$ model, relevant for high $T_c$
superconductors. This procedure yields a decomposition of the electron 
field into a product of two ``semionic" fields, i.e. fields obeying 
abelian braid statistics with statistics parameter $\theta={1\over 4}$, 
one carrying
the charge and the other the spin degrees of freedom.
A mean field theory is then shown to reproduce correctly the large distance
behaviour of the correlation functions of the 1D $t-J$ model at $t>>J$. 
This result shows that to capture the essential physical properties of 
the model one needs a specific ``semionic" form of spin--charge separation.

\vskip 1.5truecm

$^*$ Supported in part by M.P.I. This work is carried out in the framework
of the European Community Programme ``Gauge Theories, Applied Supersymmetry
and Quantum Gravity" with a financial contribution under contract 
SC1--CT--92--D789.

\vfill\eject

\vskip 0.5truecm

{\bf 1.\ Introduction}

\vskip 0.3truecm

\qquad
It is widely believed that the two--dimensional (2D) $t-J$ model 
captures many essential physical properties of the $Cu-O$ planes 
characterizing a large class of high --$T_c$ superconductors.

\qquad
The hamiltonian of the model is given by

$$
H= P_G \Bigl[\sum_{<ij>} - t(c^\dagger_{i\alpha} c_{j\alpha} + h.c.)+ 
J c^\dagger_{i\alpha}  
{\vec \sigma_{\alpha\beta} \over 2} c_{i\beta} \cdot c^\dagger_{j\gamma} {\vec 
\sigma_{\gamma\delta} \over 2} c_{j\delta} \Bigr] P_G,  \eqno(1.1)
$$

where $c_{i\alpha}$ is the annihilation operator of a spin ${1\over 2}$ 
fermion (in this paper called electron)  
at site $i$ of a square lattice, corresponding in the physical 
system to a hole on the $Cu$ site, and $P_G$ is the Gutzwiller projection
eliminating double occupation, modelling the strong on--site Coulomb 
repulsion (see e.g. [1]).

\qquad
In spite of the enormous efforts made so far,  
we still do 
not have a good mean--field theory for the 2D $t-J$ model, 
i.e. such a successful saddle--point 
that we can apply the standard quantum many--body techniques to calculate the
fluctuations around it for  describing the essential physics. On the other 
hand, we have a much better understanding of the 1D $t-J$ model. In fact, 
the strictly related large $U$ Hubbard model has been solved by 
Bethe--Ansatz [2] and for $U \sim +\infty$ the ground state wave function
can be written [3] as a product of a Slater determinant for spinless 
fermions, describing the charge degrees of freedom, and a ground state wave
function for a ``squeezed Heisenberg chain", i.e. a chain where all
empty sites are "squeezed out", describing the spin degrees of freedom.
For a finite number of electrons, the pseudomomenta of the spinless
fermions are related to the ``spin rapidity" of the spin degrees freedom, 
but in the thermodynamic limit the distribution of the pseudomomenta 
becomes a constant [4], i.e., that of a free spinless fermion system.
For the large $U$ Hubbard model and for the $t-J$ model at $t=J$,
the large scale behaviour of several correlation functions have been
computed combining Luttinger--liquid [5] and
conformal field theory [6] techniques. Related results for
$t>>J$ have also been obtained
in [7], by means of a more standard quantum field-theory approach,
using a special mean field treatment of the non-local, or "string"
field  operators.
Typical contributions for an euclidean 
two--point functions are found to be of the form

$$
{e^{i {\pi \over 2} \rho n} \over (x - i v_c t)^{\alpha^-_c} (x+ i v_c 
t)^{\alpha^+_c} (x - i v_s t)^{\alpha^-_s} (x + i v_s t)^{\alpha^+_s}} 
\eqno(1.2)
$$

for suitable $n \in {\bf Z}, \alpha^\pm_c, \alpha^\pm_s \in {\bf Q}$,
where $\rho$ is the electron density, $v_c$ and $v_s$ are the charge 
and spin velocities, respectively. Hence the correlation functions 
also exhibit charge--spin separation. These results show that
the key features of the 1D model can be understood in terms of two 
low--energy excitations, the holon, charged but spinless and the spinon, 
neutral with spin ${1\over 2}$.  

\qquad
Partly on the basis of  analogy with the 1D model, Anderson conjectured 
[8] that the physics of the 2D model can also be understood in terms of 
low energy excitations, charged but spinless (holons) and neutral with spin
${1\over 2}$ (spinons). Depending on the statistics of the holons 
(sometimes called ``slave--particles") we have the slave--fermion [9],  
slave--boson [10] and the more exotic slave--semion approach, advocated 
by Laughlin [11] 
(generalized to a slave--anyon approach in [12]). The semions are special 
kind of 
anyons (see e.g. [13]), i.e. excitations obeying abelian braid statistics, with 
statistics parameter $\theta = {1\over 4}$, i.e. the exchange of the 
field operators creating such excitations produces a phase factor 
$\pm i$, instead of 
+1 or -1 characterizing bosons and fermions, respectively.

\qquad
Recently, a bosonization scheme for 
two--dimensional fermionic systems has been proposed,
 based on the introduction of an abelian 
Chern--Simons gauge field [14]. Such a scheme has been extended  to a 
non--abelian version in [15] and both versions have been applied to the 
$t-J$ model in [16] (see also [17,12]). The $U(1)$ Chern--Simons bosonization 
has been shown to correspond essentially to the slave--boson and 
slave--fermion approaches (depending on the choice of the gauge fixing [16]);
while the non--abelian $U(1) \times SU(2)$ Chern--Simons bosonization 
corresponds to the slave semion--approach.

\qquad
Although every bosonization scheme yields an exact identity between 
correlation functions of the original fermionic field and suitable
bosonic correlation functions, the  mean field approximation (MFA)
gives 
different results in different bosonization schemes. It is then natural to 
ask which one of these schemes has a better chance to describe 
correctly in MFA the physics of the model.
One expects that possible indications for the answer might be obtained from
comparison with the known analytical results of the 1D $t-J$ model.

\qquad
One is then naturally led to discuss a dimensional reduction of 
Chern--Simons bosonization to 1D systems. One can verify that this 
reduction corresponds, roughly speaking, to a Jordan--Wigner--like 
bosonization for the gauge group $U(1)$ and a suitable non--abelian 
generalization of it for the gauge group $U(1) \times SU(2)$.

\qquad
In this paper we show that the large distance behaviour of the correlation 
functions of the 1D model are indeed reproduced by a mean field theory 
of the $U(1)
\times SU(2)$ bosonization; thus, interpreting 
spinon and holon fields as the  1D counterparts of semion fields, i.e.
they obey abelian braid statistics with statistics parameter $\theta={1\over
4}$ (in 1D only the statistics of fields, but not those of excitations are 
well defined, see, e.g.,[18].) It turns out that this dimensional reduction 
gives essentially a
more systematic justification and refined structure to the approach 
followed in [7].

\qquad
This result shows how to obtain the features of the 1D model in terms of 
standard quantum field theory techniques, and encourages us to pursue the
study of the non--abelian Chern--Simons bosonization of the 2D $t-J$ model, 
suggesting some ideas for  developing  a possibly reasonable 
mean--field treatment. Furthermore, it implicitly supports the 
interpretation of spinons and holons of the 2D model as semions, if they 
are still well defined excitations.

The plan of the paper is the following: 

\smallskip

\item{--} in sect. 2 we outline the Chern--Simons bosonization scheme and apply it 
to the 2D $t-J$ model;

\smallskip

\item{--} in sect. 3 we perform the dimensional reduction to the 1D model for 
the partition function;

\smallskip

\item{--} in sect. 4 we discuss the mean--field approximation; 

\smallskip

\item{--} in sect. 5 we perform the dimensional reduction  of  
correlation functions and discuss their  mean--field treatment.

\smallskip

Some detailed computations are deferred to Appendices.

\vskip 0.5truecm

{\bf 2.\ The Chern--Simons bosonization}

\vskip 0.3truecm

\qquad
We recall the main definitions and results of Chern--Simons bosonization 
scheme applied to spin ${1\over 2}$ fermion systems in 2D [15,16].

\qquad
Let $\Psi_\alpha, \Psi^*_\alpha$ (resp. $\Phi_\alpha, \Phi_\alpha^*), 
\alpha =1,2$ be two--component Grassmann (resp. complex) fields describing
the degrees of freedom of a spin ${1\over 2}$ canonical  
non-relativistic fermion (boson) 
field operator $\hat \Psi_\alpha$ (resp. $\hat \Phi_\alpha)$.

\qquad
Consider a system of spin ${1\over 2}$ non--relativistic fermions 
interacting via an instantaneous, spin independent two--body potential, and 
in the presence of an external, minimally coupled abelian gauge field $A$. 
The classical euclidean action of the system is denoted by 
$S(\Psi,\Psi^* | A)$. Let $B$ (resp. $V$) be a $U(1)$ $\Bigl($resp. 
$SU(2) \Bigr)$ gauge field and let

$$\eqalign{
S_{c.s.} (B) & = {1\over 4 \pi i} \int d^3 x \epsilon_{\mu\nu \rho} B^\mu 
\partial^\nu B^\rho (x) \cr
S_{c.s.} (V) & = {1 \over 4\pi i} \int d^3 x
Tr \epsilon_{\mu\nu\rho} \Bigl(V^\mu 
\partial^\nu V^\rho + {2\over 3} V^\mu V^\nu V^\rho \Bigr)(x)  \cr}
\eqno(2.1)
$$

be the corresponding euclidean Chern--Simons actions. Then the following 
bosonization formulas can be derived:

\smallskip

1) the grand--canonical partition function of the fermion system is given by:

$$\eqalign{
\int & {\cal D} \Psi {\cal D} \Psi^* e^{-S(\Psi, \Psi^* | A)} \cr
{\rm a)} & = {\int {\cal D} B \int {\cal D} \Phi {\cal D} \Phi^* e^{-[S(\Phi, 
\Phi^* |A+B) + S_{c.s.} (B)]} \over \int {\cal D} B e^{-S_{c.s.} (B)}}, \cr
{\rm b)} & = {\int {\cal D} B {\cal D} V \int {\cal D} \Phi {\cal D} \Phi^* 
e^{-[S 
(\Phi, \Phi^* |A+B+V)+2 S_{c.s.} (B) + S_{c.s.}(V)]} \over \int {\cal D} B 
{\cal D} V e^{-[2 S_{c.s.} (B) + S_{c.s.} (V)]}}, \cr} \eqno(2.2)
$$

where a) corresponds to the $U(1)$--bosonization and b) corresponds to the 
\hfill\break $U(1) \times 
SU(2)$--bosonization and gauge fixings for the respective gauge symmetries 
of the
actions are understood.

\smallskip

2) Let $\gamma_x, x = (x^0, \vec x)$ denote a string connecting $x$ 
to infinity in the  $x^0$--euclidean time plane; then one can prove an identity 
between the correlation functions of $\Psi_\alpha, \Psi^*_\alpha$ 
in the fermionic theory and the correlation functions of the
non--local fields

a)

$$
\Phi_\alpha (\gamma_x | B) = e^{i \int_{\gamma_x} B} 
\Phi_\alpha (x), \quad
\Phi^*_\alpha (\gamma_x |B) = e^{-i \int_{\gamma_x} B} \Phi_\alpha^* (x) 
$$

in the $U(1)$-bosonized theory, and

b)

$$\eqalign{
\Phi_\alpha  (\gamma_x |B, V) = e^{i \int_{\gamma_x} B} (P e^{i 
\int_{\gamma_x} V})_{\alpha\beta} \Phi_\beta (x), \cr
\Phi^*_\alpha (\gamma_x |B,V) = 
\Phi^*_\beta (x) e^{-i \int_{\gamma_x}B} 
(P e^{-i \int_{\gamma_x}V})_{\beta\alpha}\cr} \eqno(2.3)
$$

in the $U(1) \times SU(2)$--bosonized theory, for spin--singlet correlation
functions. 
$P(\cdot)$ in eq. (2.3)
denotes the path--ordering, which amounts to the usual time ordering $T(\cdot)$,
when ``time" is used to parametrize the curve along 
which one integrates. (For a careful discussion of boundary conditions and 
further details, see [15,16]).

\qquad
These bosonization formulas can be derived using a Feynman--Kac 
representation of the partition function and the correlation functions, 
expressing 
them in terms of brownian paths in ${\bf R}^2$. In this representation the
only difference between fermions and bosons are minus signs related to 
permutation in the order of initial and final points of the paths.
Using the fact that the probability for two brownian paths in ${\bf R}^2$
to intersect each other
at a fixed time is zero, one can prove that the configurations of brownian
paths appearing in the Feynman--Kac formulas are braids with probability 
1, and periodic b.c. in time convert 
these braids into knots. 
Finally, the minus signs associated with 
permutations and converting bosons into fermions are obtained from 
Chern--Simons expectation values of the exponential of the gauge fields $B, 
V$ arising from their minimal coupling to the matter bosonic fields, 
integrated over the knots formed by the 
brownian path configurations, following the construction of knots 
invariant in Chern--Simons theory (see e.g.[19]).

\qquad
As this sketchy explanation suggests, one can apply the same techniques to 
lattice theories, using a lattice version of Feynman--Kac formula [20,16], 
expressing partition function and correlation functions in terms of random 
walks in the lattice ${\bf Z}^2$, retaining the Chern--Simons gauge fields 
in the continuum version, and provided the two--body potential contains a 
\underbar{hard--core} term. In fact, the probability that two random 
walks intersect at a fixed time in ${\bf Z}^2$ is not $0$, so in order for 
the random walk configurations to be braids with probability 1, one needs a
hard--core term forbidding intersections among random walks.

\qquad
As a result of this brief discussion, one can understand 
that the 
bosonization formulas can be applied to the 2D $t-J$ model because:

\smallskip

1) the Gutzwiller projection acts as a hard--core term;

2) introducing a Hubbard--Stratonovich complex gauge field $X$ one can 
rewrite the Heisenberg term as a standard kinetic term with minimal 
coupling to $X$ (view in the bosonization procedure as
introducing an external gauge field) 
plus a two--body spin--independent potential.

\qquad
In fact, the grand--canonical partition function of the $t-J$ model at 
temperature $T= k_B / \beta$ 
(where $k_B$ is the Boltzmann constant)
and chemical potential $\mu$ can be 
rewritten [21] as: 

$$
\Xi_{t-J} (\beta, \mu) = \int {\cal D} X {\cal D}
X^* {\cal D} \Psi {\cal D} \Psi^* e^{-S_{t-J} (\Psi, \Psi^*, X, 
X^*)} \eqno(2.4)
$$

with

$$
S_{t-J} (\Psi, \Psi^*, X, X^*) =
\int^\beta_0 d \tau \sum_{<ij>} {2 \over J} X^*_{<ij>} X_{<ij>} + 
[(-t + X_{<ij>}) \Psi^*_{i \alpha} \Psi_{j \alpha}
$$
$$
+ h.c.]+ \sum_i \Psi^*_{i\alpha} ({\partial \over \partial \tau} + \mu) 
\Psi_{i\alpha} 
+ \sum_{i,j} u_{i,j} \Psi^*_{i\alpha} \Psi^*_{j\beta} \Psi_{j\beta}
\Psi_{i\alpha}, \eqno(2.5)
$$

where the two--body potential is given by

$$
u_{i,j} =\cases{+ \infty  \quad i=j \cr
-{J \over 4} \quad i,j \ {\rm n.n.} \cr
0 \quad {\rm otherwise}. \cr} \eqno(2.6)
$$

(The euclidean--time $(\tau)$ dependence of the fields here and after
 is not explicitly written, and repeated spin indices are summed 
over.)

\qquad
The bosonized action is obtained via substituting the time derivative by the 
covariant time derivative and the spatial lattice derivative by the 
covariant spatial lattice derivative: e.g. in the $U(1) \times SU(2)$
bosonization

$$
\Psi^*_{j\alpha} {\partial \over \partial\tau} \Psi_{j\alpha} \longrightarrow 
\Phi^*_{j\alpha} \Bigl[\Bigl({\partial \over \partial \tau} + i B_0 (j)
\Bigr) \unity + i  V_0 (j) \Bigr]_{\alpha\beta} \Phi_{j\beta},
$$
$$
\Psi^*_{i\alpha} \Psi_{j\alpha} \longrightarrow \Phi^*_{i\alpha}
e^{i \int_{<ij>} B} (P e^{i \int_{<ij>} V})_{\alpha\beta} 
\Phi_{j\beta}. \eqno(2.7)
$$

One can then decompose

$$
\Phi_{j\alpha} = \tilde E_j \Sigma_{j\alpha}, \eqno(2.8)
$$

with the constraint

$$
\Sigma^*_{j\alpha} 
\Sigma_{j\alpha} =1, \eqno(2.9)
$$

where $\Sigma_\alpha$ and $\tilde E$ are a 2--component and a 1--component 
complex 
lattice fields, respectively. However, such decomposition is ambiguous, 
since we can still perform a 
local $U(1)$--gauge tranformation leaving  $\Phi$ invariant:

$$\eqalign{
\tilde E_j & \rightarrow \tilde E_j e^{i \Lambda_j}, \cr
\Sigma_{j \alpha} & \rightarrow \Sigma_{j\alpha} e^{-i \Lambda_j}, \quad
\Lambda_j \in [0, 2\pi).  \cr}\eqno(2.10)
$$

Therefore, the theory expressed in terms of $\Phi$ is equivalent to the 
theory  in terms of $E$ and $\Sigma$ only with a gauge--fixing term of (2.10), 
not breaking the $U(1) \times SU(2)$ gauge invariance of the $\Phi$--theory.
As an example one can choose a Coulomb gauge for the $U(1) \times S 
U(2)$--gauge invariant field ``arg $(\Sigma_i^* P  
e^{i \int_{<ij>} V} \Sigma_j)$" supported on links:

$$
\sum_{\scriptstyle <ij>: \ell \in <ij>} \ \arg \ (\Sigma^*_i P 
e^{i \int_{<ij>} V} \Sigma_j)=0. \eqno(2.11)
$$

In terms of $\tilde E$ and $\Sigma$ the $U(1) \times SU(2)$--bosonized action
of the $t-J$ model is given by

$$\eqalign{
S_{t-J}& (\tilde E, \tilde E^*, \Sigma, \Sigma^* , X, X^*, B,V)=
\int_0^\beta d \tau \sum_{<ij>} {2\over J} X^*_{<ij>} X_{<ij>} \cr
+ & [(-t + X_{<ij>}) \tilde E_j^* e^{i \int_{<ij>}B} \tilde E_i
\Sigma^*_{i\alpha} (P e^{i \int_{<ij>} V})_{\alpha\beta} \Sigma_{j \beta} + 
h.c.] \cr
+ & \sum_j \tilde E^*_j ({\partial \over \partial \tau} -
i B_0 (j) + \mu +{J \over 2}) \tilde E_j + \tilde E^*_j \tilde E_j 
\Sigma^*_{j\alpha}
\Bigl({\partial \over \partial \tau} \unity + i 
V_0 (j) 
\Bigr)_{\alpha\beta} \Sigma_{j\beta}\cr
+ & \sum_{i,j} u_{i,j} \tilde 
E_i^* \tilde E_i \tilde E_j^* \tilde E_j + 2 
S_{c.s.} (B) + S_{c.s.} (V) \cr} \eqno(2.12)
$$

with constraint (2.9) and gauge fixings understood.

\qquad
One can easily convert $\tilde E$ into a fermion field $E$, by inverting 
the sign of $S_{c.s.} (B)$; one then omits the hard--core term in $u_{i,j}$.
(In fact to perform this transformation we couple $E$ to a new $U(1)$ gauge 
field $B'$ with action $S_{c.s.} (B')$, changing variable $B \rightarrow 
B'' = B+B'$ and integrating out $B$ the claimed result follows; one then 
rewrites $B''$ again as $B$.)
Integrating out $X, X^*$ and using the anti--commutation 
properties of the field $E$, one obtains the action

$$\eqalign{
S_{t-J} & (E, E^*, \Sigma, \Sigma^*, B, V) 
= \int^\beta_0 d\tau  \sum_j E^*_j ({\partial \over \partial\tau} + i B_0
(j) + \mu + {J \over 2} ) E_j \cr 
+E^*_j & E_j  \Sigma^*_{j\alpha} \Bigl({\partial \over \partial \tau} 
\unity + i V_0 (j) \Bigr)_{\alpha\beta} \Sigma_{j\beta} \cr
+ \sum_{<ij>} & (-t E^*_j e^{i\int_{<ij>} B} E_i \Sigma^*_{j\alpha}
(P e^{i \int_{<ij>} V})_{\alpha\beta} \Sigma_{i\beta} + h.c.)  \cr
+ {J \over 2} & E^*_j E_j E^*_i E_i \Bigl\{| \Sigma_{i\alpha}^* (P e^{i 
\int_{<ij>} V})_{\alpha\beta}\Sigma_{j\beta} |^2 - {1\over 2} \Bigr\} 
- 2 S_{c.s.} (B) + S_{c.s.} (V). \cr} \eqno(2.13)
$$

Notice that if instead of $E$ we use the bose field $\tilde E$, the $J$ 
term gets an opposite sign due to the commutation properties of $\tilde E$.

\qquad
In terms of these variables the correlation functions of $\Psi_\alpha$ are 
given in the bosonized theory by

$$
\Phi_\alpha (\gamma_x | B, V)= e^{i \int_{\gamma_x} B} 
E_x (P e^{i \int_{\gamma_ x} V} )_{\alpha\beta} \Sigma_{x \beta}. \eqno(2.14)
$$

Therefore one can view our original electron field as a product of $U(1) \times 
SU(2)$--gauge invariant fields $E(\gamma_x |B) \equiv e^{i \int_{\gamma_x}
B} E_x$ (holon) and  $\Sigma_\alpha (\gamma_x |V) = \hfill\break
(P e^{i \int_{\gamma_x} 
V})_{\alpha\beta} \Sigma_{x \beta}$ (spinon).
{}From the coefficient of the Chern--Simons terms of $B$ and 
$V$ one can derive [15] that the corresponding 
non local holon and spinon field operators 
obey semionic statistics, i.e. their 
statistics parameter is $\theta = {1\over 4}$.

\vskip 0.5truecm
{\bf 3.\ Reduction to 1D: the partition function}
\vskip 0.3truecm

\qquad
The reduction of the 2D--system to 1D is obtained by letting the electron
field sit on the lattice 
${\bf Z}$  instead of ${\bf Z}^2$.  
Denote by 1  the spatial dimension in ${\bf R}^2$ along the 1D lattice
{\bf Z}.

\qquad
Since in the partition function the only dependence of $B_2$ and $V_2$ is 
in the Chern--Simons action, one can integrate them out yielding the 
constraints

i)

$$
\epsilon_{\mu\nu} \partial^\mu B^\nu =0,
$$

ii) 

$$
\epsilon_{\mu\nu} (\partial^\mu V^\nu + V^\mu V^\nu)=0 \quad ; \quad 
\mu,\nu = 0,1. \eqno(3.1)
$$
The Coulomb gauge--fixing condition for the $U(1)$ gauge invariance, 
$B_1 = 0$, with the boundary condition $B_\mu (x^1 = + \infty) =0,$
together with the constraint i) yields $B_\mu =0$.

\qquad
The constraint ii) is solved by

$$
i V_\mu (x) = g^\dagger (x) \partial_\mu g(x), \qquad g(x) \in 
SU(2) \eqno(3.2)
$$

and in terms of $g$ and $\Sigma$ the $SU(2)$ gauge transformation reads as

$$
g_j \rightarrow r_j g_j, \qquad \Sigma_j \rightarrow r_j \Sigma_j,
\Sigma_j^* \rightarrow \Sigma^*_j r^\dagger_j, \eqno(3.3)
$$

where $r_j \in SU(2)$ and $g_j (\tau) \equiv g(\tau, j)$.

\qquad
To gauge--fix (3.3) we use a ferromagnetic reference spin configuration,
by setting

$$
\Sigma_j =  \Bigl(\matrix {1 \cr
0 \cr} \Bigr). \eqno(3.4)
$$

Finally the gauge--fixing (2.11) reads, being reduced to 1D:

$$
\arg\ (\Sigma^*_i g_i^\dagger g_j \Sigma_j) = \delta. \eqno(3.5)
$$

Choosing the constant $\delta$ in (3.5) to be $0$ for later 
convenience, in the gauge (3.4), this implies that ($g^\dagger_i g_j)_{11}$
is real and positive, hence $g^\dagger_i g_j$ is in the
linear span of $\unity, \sigma_x, \sigma_y$, because $g^\dagger_i g_j \in
SU(2)$.

\qquad
Let us count the degrees of freedom  (d.f.): the matter fields $E, 
E^*, \Sigma_\alpha, \Sigma^*_\alpha$ have $2+(4-1)$ d.f., 
the field $g$ has 3 
d.f. and the gauge fixings (3.5) and (3.4) eliminate 1+3 d.f., so  we are 
left exactly  with 4 d.f. as we had for $\Psi_\alpha, \Psi^*_\alpha$; the 
charge degrees of freedom are carried by $E, E^*$ , while the spin degrees
of freedom are carried by $g$.

The partition function of the 1D $t-J$ model can be rewritten as

$$
\Xi_{t-J} = \int {\cal D} g {\cal D} E {\cal D} E^* e^{- S_{t-J} (E, E^*, 
g)}
\prod_{<ij>} \delta (\arg\ (g^\dagger_i g_j)_{11}) \eqno(3.6)
$$

with

$$\eqalign{
S_{t-J} & = \int^\beta_0 d \tau \sum_j E^*_j \Bigl({\partial \over 
\partial \tau} + \mu + {J\over 2} + (g^\dagger_j \partial_0 g_j)_{11} 
\Bigr) E_j \cr
+ \sum _{<ij>} & (-t E^*_j E_i (g^\dagger_j g_i)_{11} + h.c.)
+ {J \over 2} E^*_j E_j E^*_i E_i\Bigl\{ |(g^\dagger_j g_i)_{11}|^2 
- {1\over 2}\Bigr\}.\cr} \eqno(3.7)
$$

[We used $\int_{S^1 \times {\bf R}^2} (g^{-1} dg)^3 =0]$.

\qquad
We now find a configuration, $g^m (E, E^*)$ of $g$ minimizing the action 
for a fixed configuration of the holons described by $E$. The idea behind 
is that we can then treat spin fluctuations around $g^m (E, E^*)$ in some 
MFA.
To find $g^m (E, E^*)$ we use the ``Feynman--Kac" random--walk
representation for the $E$--path integration. It can be considered 
as a lattice version [16, 20] of the representation of a 
Feynman path--integration 
over non--relativistic fermion fields in terms of path--integrals over 
trajectories of a fixed number of Fermi particles summed over all 
possible number of particles (see e.g.[22]).

\qquad
More precisely, let $\Delta$ denote the 1D lattice laplacian defined on 
a scalar lattice field $f$ by 

$$
(\Delta f)_i = f_{i+1} + f_{i-1} - 2 f_i,
$$

let $d\mu (\omega)$ denote the measure on the random walks $\omega$ on the 
1D lattice such that

$$
(e^{-\beta \Delta})_{ij} = \int_{\scriptstyle \omega (0) =i \atop
\scriptstyle \omega (\beta)=j} d\mu (\omega), \quad \beta> 0, \eqno(3.8)
$$

let $P_N$ be the group of permutations of $N$ 
elements and for $\pi \in P_N$ let $\sigma (\pi)$ denote the order of 
the permutation. Then the partition function of the $E$--system in a
fixed $g$ configuration can be written as:

$$
Z (g) \equiv \int {\cal D} E {\cal D} E^* e^{- S_{t-J} (E, E^*, g)} =
$$
$$
= \sum^\infty_{N=0} {e^{\beta (\mu+ {J\over 2})N} \over N!} 
\sum_{\pi\in P_N} (-1)^{\sigma(\pi)} \sum_{j_1, ..., j_N} 
\int_{\scriptstyle \omega_r (0)= j_r \atop \scriptstyle \omega_r (\beta) 
=j_{\pi(r)}} \prod^N_{r=1} d\mu (\omega_r) \cdot
$$
$$
\Bigl[\prod_{r=1}^N
\prod_{\scriptstyle <ij> \in \omega_r^\bot} t (g^\dagger_i g_j)_{11} 
\exp \ \Bigl\{\int_{\omega^{\|}_r} d\tau
(g^\dagger \partial_\tau g)_{11}  \Bigr\}\cdot 
$$
$$
\exp\ \Bigl\{{-\sum^N_{r,j=1}} {J\over 2} \int^\beta_0 d\tau 
\delta_{|\omega_r (\tau)
- \omega_j(\tau)|,1} [|(g^\dagger (\omega_{r(\tau)}) g (\omega_j 
(\tau)))_{11}|^2 - {1\over 2}] \Bigr\} \Bigr], \eqno(3.9)
$$

where $\omega^\bot_r$ (resp. $\omega_r^\|)$ denotes the component of 
$\omega_r$ 
perpendicular (resp. parallel) to the time axis.
We first notice that, due to the Pauli principle in 1D,  
only the trivial permutation contributes to (3.9), 
since the random walks cannot intersect each other.

\qquad
Consider a fixed configuration of random walks $\underline{\omega} 
= \{\omega_1, ... \omega_N \}$; using the inequality 
$|(g^\dagger_i g_j)_{11}| \leq 1$ and $Re 
(g^\dagger 
\partial_\tau g)_{11}=0$, one can bound from above the absolute value of its
contribution to (3.9) in square bracket  by 

$$
\prod^N_{r=1} \prod_{\scriptstyle <ij>\in \omega_r^\bot} t \ 
\exp\  \Bigl\{ {-\sum_{\scriptstyle r,j =1}^N} \{{J\over 2} \int^\beta_0
d\tau 
\delta_{|\omega_r(\tau) - \omega_j (\tau)|,1} \cdot [-{1\over 2}]\} \Bigr\}.
$$

This upper bound is exactly saturated by a configuration
of $g$ satisfying 

$$
(g^\dagger \partial_\tau g)_{11} =0  \qquad {\rm on}\quad \omega_r^\|,
\eqno(3.10)
$$

$$
(g^\dagger_i g_j)_{11} = e^{i \delta_{<ij>}}, \quad \delta_{<ij>} = - 
\delta_{<ji>}, \quad {\rm if} <ij> \in \omega_r^\bot, \eqno(3.11)
$$

$$
\Bigl(g^\dagger (\omega_r (\tau) g(\omega_j (\tau) \Bigr)_{11} =0 \qquad
{\rm if} \quad |\omega_r (\tau) - \omega_j (\tau)|=1. \eqno(3.12)
$$

(The arbitrariness in $\delta_{<ij>}$, independent on time, follows from

$$
\sum_{\scriptstyle <ij> \in \omega_r^\bot} \delta_{<ij>} =0, 
$$

by periodicity in the time direction.)

\qquad
If we further want to satisfy on $\omega_r^\bot$ the constraint

$$
\arg\ (g^\dagger_i g_j)_{11} = 0, \eqno(3.13)
$$

we must set $\delta_{<ij>} = 0$.

\qquad
To sum up, we obtain the following minimizing configuration  $g^m \equiv g^m 
(\underline{\omega})$:   
 we choose $g^m$ constant during the period when the particle 
does not jump, to satisfy (3.10);  while in a link $<ij>$ using the 
representation

$$
g^{m\dagger}_j g_i^m = \cos\ \theta \unity + i {\rm sen} \theta
\vec \sigma \cdot \vec n, \eqno(3.14)
$$

we see that (3.12) is satisfied by choosing $\theta = {\pi \over 2}, n_z =0$
on links not in $\underline{\omega}^\bot$, whereas  for links in 
$\underline{\omega}^\bot$ we 
choose $\theta = 0$ to satisfy (3.11). Therefore, we can represent the 
minimizing configuration as 

$$
g^m_j (\tau | \underline{\omega})= e^{i {\pi \over 2} \sigma_x \sum_{\ell < 
j} \sum^N_{r=1} \delta_{\omega_r (\tau), \ell}}. \eqno(3.15)
$$

(At $\theta = {\pi\over 2}$, different choices of $n_x, n_y$, but
satisfying the condition 
$n^2_x + n^2_y =1$, give the same element of $SU(2)$ in (3.14);
we choose to work with $n_x =1, n_y =0$.)

\qquad
Reexpressed in terms of the fields $E, E^*$ and denoted by $g^m (E,
E^*)$ the minimizing configuration (3.15) becomes

$$
g^m_j (\tau |E, E^*)= e^{i{\pi \over 2}\sigma_x \sum_{\ell <j} E^*_\ell
E_\ell (\tau)}.\eqno(3.16)
$$

(We used here the left--continuity of the paths $\omega$ [20], so that
at a jumping time $\omega_r (\tau) = \lim\limits_{\epsilon \searrow 0} \omega_r 
(\tau + \epsilon)$, and  the sign $\lim\limits_{\epsilon \searrow 0}$ 
means $\epsilon \to 0 $ from the positive side.)

It is natural to perform first the integration over $g$ by changing 
variable from  
the $SU(2)$--valued field $g$ to an $SU(2)$--valued field $U$ describing 
fluctuations around the minimizing configuration $g^m$ and defined by

$$
g = U g^m. \eqno(3.17)
$$

For $j \in {\bf Z}$ we set

$$
[j] = \cases{ 1 & \quad $j$ {\rm odd}, \cr
2 & \quad $j$ {\rm  even}, \cr}\eqno(3.18)
$$

and 

$$
\tilde j (\tau |\underline{\omega})
= \sum_{\ell < j} \sum^N_{r=1} \delta_{\omega_r 
(\tau), \ell}. \eqno(3.19)
$$

To simplify the notation with the meaning being  clear from the context, we 
set $\tilde j (\tau | \underline{\omega}) = \tilde j (\tau)$, so that, e.g.,
the r.h.s. of eq.(3.15) reads $e^{i {\pi \over 2} \sigma_x \tilde j (\tau)}$.
Then, in terms of $U$ and the random walks $\underline{\omega}$, 
the partition function is given by

$$
\Xi_{t-J} = \sum^\infty_{N=0} {e^{\beta (\mu + {J\over 2})N} \over N!}
\sum_{j_1, ..., j_N} \int_{\scriptstyle \omega_r(0) = j_r \atop \scriptstyle 
\omega_r (\beta)= 
j_r} \prod^N_{r=1} d\mu (\omega_r) 
\int {\cal D} U 
$$
$$
\prod_{\scriptstyle <ij>\in \omega^\bot_r} \delta
\Bigl(\arg \ (U^\dagger_i U_j)_{[\tilde i][\tilde j]} \Bigr)\prod_{\scriptstyle <ij> \in \omega^\bot_r} t (U_i^\dagger U_j)_{[\tilde 
i][\tilde j]} \exp\ {\int_{\omega_r^\|} 
d\tau(U^\dagger \partial_\tau U)_
{[\tilde \omega_r(\tau)][\tilde \omega_r(\tau)]}}\cdot
$$
$$
\exp\ \Bigl\{{-\sum^N_{j, r=1} {J\over 2} \int^\beta_0 d\tau \ 
\delta_{|\omega_r (\tau) 
- \omega_j (\tau)|,1}} [|(U^\dagger (\omega_r(\tau)) U (\omega_j (\tau))
)_{[\tilde
\omega_r (\tau)][\tilde \omega_j(\tau)]} |^2 - {1\over 2}] \Bigr\}. \eqno(3.20)
$$

Analogously, defining

$$
\tilde j (\tau | E, E^*)= \sum_{\ell <j} 
E^*_\ell E_\ell (\tau), \eqno(3.21)
$$

we have in terms of $E, E^*, U$:

$$
\Xi_{t-J} = \int {\cal D} E {\cal D} E^* \int {\cal D} U e^{-S_{t-J} (U,
E, E^*)} \prod_{<ij>} \delta (\arg\ (U^\dagger_j U_1)_{[\tilde 
j][\tilde i]} \eqno(3.22)
$$

with 

$$
S_{t-J} (U, E, E^*)= \int_0^\beta d\tau \sum_j E^*_j ({\partial \over 
\partial \tau} + \mu + {J \over 2} +(U^\dagger \partial_\tau U)_{[\tilde j]
[\tilde j]}) E_j +
$$
$$
\sum_{<ij>} \Bigl[(-t E^*_j E_i) (U^\dagger_j U_i)_{[\tilde j][\tilde i]} +
h.c. \Bigr] + {J \over 2} E^*_j E_j E^*_i E_i \{| (U^\dagger_j 
U_i)_{[\tilde j][\tilde i]}|^2 - {1\over 2}\}. \eqno(3.23)
$$

Notice that up to now no approximations have been made and (3.20), 
(3.22--23) are 
exact rewritings  of the partition functions of the 1D $t-J$ model. 

\vskip 0.5truecm
{\bf 4.\ Mean field approximation}
\vskip 0.3truecm

\qquad
We now wish to compare our result (3.20) with the Bethe--Ansatz ground state 
wave function of the $U \sim + \infty$ Hubbard model, essentially equivalent to
the $t-J$ model at $J \sim + 0$.
To make the comparison we first restrict ourselves to $T \sim 0$, finite 
volume $V$ and finite number of electrons $N$, with $V, N$ being large and we 
assume $\rho \equiv N/V = 1 - \delta, \delta << 1$ and  $t >>J$.

\qquad
Then we carry out the following mean--field treatment: 

\smallskip

1) We assume that the spin fluctuations can be treated in mean field in the
hopping term of the charged particles and we denote by $t_R$ 
the renormalized hopping;

\smallskip

2) Since the motion of the charged particles is much faster than the spin 
motion, we replace $\omega_r (\tau)$ by its average in time, and since 
the paths cannot overlap \hfill\break 
$< \omega_r (\tau) > = {r \over 1-\delta}$.
So we replace for the spin motion the original  chain by a ``squeezed chain"
of lattice spacing  $(1-\delta)^{-1}$ and accordingly we replace 
\hfill\break
$\delta_{|\omega_r (\tau) - \omega_j (\tau)|,1}$ by its mean value 
$(1-\delta) \delta_{|r - j|, 1}$ in the ``squeezed chain". The 
corresponding renormalized spin coupling constant is denoted by $J_R$.

\qquad
After making these approximations the canonical partition function
decouples into 
a product of the  partition function for  a free charged holon system on the 
original lattice and the partition function 
for a spin ${1\over 2}$ quantum Heisenberg chain on 
the ``squeezed lattice" with lattice spacing $(1-\delta)^{-1}$:

$$
Z_{M.F.} (N) = \int {\cal D} E {\cal D} E^* e^{-\int d\tau 
\sum_i [E_i^* \partial_\tau E_i -
\sum_{<ij>} t_R (E^*_i E_j + h.c.)}
$$
$$
\delta(\sum_i E^*_i E_i - N) \cdot \int^s {\cal D} U \delta (\arg
(U^\dagger_i U_j)_{[i][j]})
$$
$$
e^{-\int d\tau \sum_j \{(U^\dagger_j \partial_\tau U_j)_{[j][j]} +
\sum_{<ij>} {J_R \over 2} \{|(U_i^\dagger U_j)_{[i][j]} |^2 
- {1\over 2} \},} \eqno(4.1)
$$

where $\int^s$ means that the variables integrated over belong to the 
``squeezed lattice". (The restriction to the lattice of finite volume 
$V$ is understood.) Therefore, this MFA correctly reproduces 
the features of the Bethe--Ansatz ground state
wave  function outlined in the Introduction.

\qquad
After making these MFAs one can then take again the 
thermodynamic limit of the system with a fixed density of holes.
Correlation functions of fields will be discussed in the next section in 
this limit.

\qquad
To verify that the $U$--system in (4.1) is actually the quantum  
Heisenberg chain one can use the $CP^1$ representation of $U$:

$$
U_j =  \Bigl(\matrix{b_{j \uparrow} & -b^*_{j\downarrow} \cr
b_{j\downarrow} & b^*_{j \uparrow} \cr} \Bigr)
\equiv \Bigl(\matrix{b_{j1} & -b^*_{j2} \cr
b_{j2} & b^*_{j1} \cr} \Bigr), \eqno(4.2)
$$

where $b_\alpha$ is a complex 2--component field constrained by

$$
b^*_{j\alpha} b_{j \alpha} =1. \eqno(4.3)
$$

In fact, setting

$$
{\tilde b_{j1} \choose \tilde b_{j2}} = U_j  e^{i {\pi\over 2}
\sigma_x j} {1\choose 
0}, \eqno(4.4)
$$

the $U$--action becomes

$$
S(\tilde b, \tilde b^*) = \int d\tau \sum_i \tilde b^*_{i \alpha} 
\partial_\tau \tilde b_{i \alpha} + {J_R \over 2} \sum_{<ij>} \tilde 
b^*_{i\alpha} \vec \sigma_{\alpha\beta} \tilde b_{i \beta} \cdot
\tilde b^*_{j\gamma} \vec \sigma_{\gamma\delta} \tilde b_{j \delta}.
\eqno(4.5)
$$

One recognizes in (4.5) the action of the quantum Heisenberg chain in the 
Schwinger boson representation.
The gauge--fixing condition for $U$ becomes then

$$
\arg\ \tilde b_{i\alpha}^* \tilde b_{j\alpha} =0, \eqno(4.6)
$$

and it can be seen as a gauge--fixing condition for the $U(1)$--gauge 
transformation

$$
\tilde b_{j\alpha} \rightarrow \tilde b_{j\alpha} e^{i \zeta_j} \quad, 
\quad
\tilde b_{j\alpha}^* \rightarrow \tilde b^*_{j\alpha} e^{-i \zeta_j} \quad,
\quad \zeta_j \in [0, 2\pi). \eqno(4.7)
$$

(Notice that the invariance of the term with time derivative is guaranteed by
the periodic b.c. in time  and the constraint (4.6).)

\qquad
One can go a step further by using the identity $\vec 
\sigma_{\alpha\beta} \cdot \vec \sigma_{\gamma\delta} = 2 
\delta_{\alpha\delta} 
\delta_{\beta\gamma} - \delta_{\alpha\beta} \delta_{\alpha\delta}$ and then
treating the quartic $\tilde b$ term in the Gor'kov decoupling approximation.
 Assuming 
translational invariance and taking into account the $\tilde b$--gauge 
fixing, this yields 

$$
\tilde S_{MF} (\tilde b^*, \tilde b)= \int d\tau \sum_j \tilde 
b_{j\alpha}^* \partial_\tau \tilde b_{j\alpha} + \sum_{<ij>} \tilde J_R 
(\tilde b_{j\alpha}^* \tilde b_{i\alpha} + h.c.), \eqno(4.8)
$$

where $\tilde J_R = <\tilde b_{i\alpha}^* \tilde 
b_{j\alpha} > J_R$.

\qquad
In the Appendix A  
we show that this system is equivalent to
a system of spin ${1\over 2}$ free fermions, described
by two component Grassmann fields $f_\alpha, f^*_\alpha$,
Gutzwiller projected and the 
large--scale properties of the quantum Heisenberg chain are exactly 
reproduced by applying the standard abelian bosonization 
to this mean--field theory. The bosonization is performed
in terms of a real scalar field $\phi_-$ with (euclidean) Thirring--Luttinger
action 

$$
S (\phi_-) = {1 \over 4\pi} \int [(\partial_0 \phi_-)^2 + v^2_s (\partial_1 
\phi_-)] \eqno(4.9)
$$

and with the following bosonization formulas for the left and right movers 
of the Gutzwiller projected fermions:

$$\eqalign{
f_1^R (x) = & f^{*R}_2 (x) \sim (2\pi)^{-{1\over 4}}
D_- (x, {1\over 2}): e^{-{i\over 2} 
\phi_- (x)}:, \cr
f^R_2 (x) = & f^{*R}_1 (x) \sim (2\pi)^{-{1\over 4}}
D_- (x, -{1\over 2}) : e^{{i\over 2} \phi_- 
(x)}:, \cr
f^L_1 (x) = & f^{*L}_2 (x) \sim (2\pi)^{-{1\over 4}}
D_- (x, {1\over 2}) : e^{{i\over 2} 
\phi_-(x)}:, \cr
f^L_2 (x) = & f^{*L}_1 (x) \sim (2\pi)^{-{1\over 4}}
D_- (x, -{1\over 2}): e^{-{i\over 2} 
\phi_(x)}:, \cr} \eqno(4.10)
$$

where $v_s \sim \tilde J_R$ is the spin velocity and
$D_- (x, \pm {1\over 2})$ is a disorder field,
see [23] and Appendix A.

\qquad
An interesting consequence of eq.(4.10)
and (A.11) is that the spin ${1\over 2}$ Gutzwiller 
projected fermion operator $\hat f$ reconstructed from the (euclidean) field
$f$ does not obey fermionic commutation relations but rather``semionic" 
with statistics parameter $\theta = {1\over 
4}$ .
In fact, let $x^\epsilon_\pm = (\epsilon, x^1), x^1 \mma 0$, let ${\cal 
F}_+ ({\cal F}_-) $ be a polynomial of exponentials of $f$ and disorder 
fields with support in $x^0 > \delta > \epsilon \ (x^0 < -\delta 
<-\epsilon, {\rm resp.})$, then (see Appendix A)

$$\eqalign{
& \lim_{\epsilon \searrow 0} \langle {\cal F}_- f_1^R (0) f^R_1 
(x^\epsilon_\pm) 
{\cal F}_+ \rangle = 
\lim_{\epsilon \searrow 0} e^{-{i\over 4} [\arg\ (x^\epsilon_\pm) + 
\arg\ (- x^\epsilon_\pm)]} e^{{i \over 4} [\arg\ (x^{-\epsilon}_\pm) 
+ \arg\ (x^{-\epsilon}_\pm)]} \cdot \cr
& \langle {\cal F}_- f^R_1 (0) f^R_1 (x^{-\epsilon}_\pm) {\cal F}_+ \rangle =
e^{\mp {1\over 4} \pi} e^{\pm {i\over 4} 3\pi} \lim_{\epsilon \searrow 
0} \langle {\cal F}_- f^R (0) f^R (x^{-\epsilon}_\pm) {\cal F}_+ \rangle \cr}
\eqno(4.11)
$$

and similar results for $f^L$.
A standard result (see e.g. [24]) of axiomatic quantum field theory 
then gives the equal-time commutation relations

$$
\hat f (x^1) \hat f(y^1) = e^{\pm {i\pi \over 2}} \hat f (y^1) \hat f 
(x^1), \qquad x^1  \mma y^1.\eqno(4.12)
$$

\vskip 0.5truecm
{\bf 5.\ Reduction to 1D: correlation functions}
\vskip 0.3truecm

\qquad
In this section we discuss the dimensional reduction to 1D  for  the 
correlation functions of the $U(1) \times SU(2)$ bosonized $t-J$ model. 
In the MFA discussed in sect. 4 one obtains exactly 
the large--scale behaviour of the correlation functions of the 1D 
$t-J$-model as derived in [5,6] by means of Luttinger liquid 
and conformal field-theory techniques. 
Furthermore, a simple interpretation of these results  emerges in terms 
of two  elementary excitations, the charged spinless holon and the spin 
${1\over 2}$ neutral spinon: the electron field operator can be decomposed 
into a product of non local holon and spinon field operator, 
both obeying abelian 
braid statistics with statistics parameter $\theta = {1\over 4}$.
Therefore, it is natural to view (althought not compulsory due to the 
ambiguity of excitation statistics appearing in 1D)  both 
 holons and spinons as 1D analogues of semions, or, using a language 
more accepted in 1D (Zamolodchikov)--parafermions of order 4 [25].

\qquad
We start by noticing that one should choose carefully the curve $\gamma_x$, 
needed 
in (2.14) to define the bosonized electron field, in order to have a good 
dimensional reduction of correlation functions. In fact as mentioned in 
sect. 2, Chern--Simons bosonization is well defined only if there are no 
intersections in the paths on which the Chern--Simons gauge fields are 
integrated over. This was automatically ensured by the Gutzwiller 
projection for the random walks representing the virtual worldlines 
of particles, appearing in the partition function, but if we choose e.g. 
$\gamma_x$ as a straight line in the 1--direction $\gamma_x$ may intersect 
many of these worldlines making the bosonization procedure ill--defined.
This can be avoided by choosing $\gamma_x$ as the path in ${\bf R}^3$ 
given by the union of the straight line going from $x= (x^0, x^1, x^2=0)$
to $x^\epsilon = (x^0, x^1, x^2 = \epsilon)$ and the straight line joining 
$x^\epsilon$ to $-\infty$ in the $1$--direction. One then takes the limit 
$\epsilon \searrow 0$.

\qquad
We discuss a general bosonization formula for fermion fields $\Psi_\alpha, 
\Psi^*_\alpha$, but one should keep in mind that our formulas 
apply only if the fermion indices appearing in the expectation values 
are saturated in a spin--singlet combination, since the non--abelian
Chern--Simons bosonization formulas have been proved only for 
spin--singlet correlations[16]. Of course, we may use the global $SU(2)$ 
invariance of the $t-J$ model action to reduce more general correlations 
to linear combinations of spin--singlet correlations,
e.g., $\langle \Psi^*_{x^1 \mu} (x^0) \Psi_{y^1 \nu} (y^0) \rangle =
{\delta_{\mu\nu} \over 2} \langle \Psi^*_{x^1 \alpha} (x^0) \Psi_{y^1 \alpha} 
(y^0) \rangle$.

\qquad
With a spin--singlet arrangement of indices understood and assuming $x^0_r 
< x^0_{r+1} < y^0_s <y^0_{s+1}$, $r, s=1,..., n-1$, the non--vanishing 
$2n$--point fermion correlation functions are given in terms of $E, B, \Sigma, 
V$, by

$$\eqalign{
\langle & \prod_{r=1}^n \Psi^*_{x^1_r \alpha_r} (x^0_r) \prod^n_{s=1} 
\Psi_{y^1_s \alpha_s} (y^0_s) \rangle = \cr
& = \langle \prod^n_{r=1} E^*_{x^1_r} (x^0_r) \exp \ 
\{-i \int_{\gamma_{x_r}} 
B \} [\Sigma^*_{x^1_r} (x^0_r) (P \exp \  \{-i \int_{\gamma_{x_r}} V \}) 
]_{\alpha_r} \cr
& \prod^n_{s=1} E_{y^1_s} (y^0_s) \exp \{i \int_{\gamma_{y_s}} B\} 
[(P \exp \{i 
\int_{\gamma_{y_s}} V \}) \Sigma_{y^1_s} (y^0_s)]_{\beta_s} \rangle. \cr}
\eqno(5.1)
$$

The dependence of $B_2$ and $V_2$ is now both in the Chern--Simons action 
and in the strings $\{\gamma_{x_r}, \gamma_{y_s}\}$ attached to the fields, 
in the part involving the infinitesimal excursion in the 2--direction, as 
specified above.

\qquad
Let us first discuss the $B_2$--integration which is simpler. It yields the 
constraint 

$$
\epsilon_{\mu\nu} \partial^\mu B^\nu (z) = \pi \chi_{[0, \epsilon]} (z^2)
\{\sum_s \delta (z^1 - y_s^1)
\delta (z^0 - y^0_s) - \sum_r \delta (z^1 - x^1_r) \delta (z^0 - x^0_r) 
\} \eqno(5.2)
$$

solved, with the Coulomb gauge fixing $B_1 =0$ and b.c. $B_\mu (x^1 =
+ \infty) =0$ by

$$
B_0 (z) = \pi \chi_{[0, \epsilon]} (z^2) \{\sum_s \delta (z^0 - y^0_s) 
\Theta (y^1_s - z^1) - \sum_r \delta (z^0 - x^0_r) \Theta (x^1_r - z^1) \},
\eqno(5.3)
$$

where $\chi_{[a,b]}$ denotes the characteristic function of the interval 
$[a,b]$ and $\Theta$ is the Heaviside step function.
Evaluated with a regularization procedure, the term containing $B_0$ 
in the action is given, in the presence of the field insertions, by

$$
-\sum_r i {\pi \over 2} \sum_{\ell < x^1_r} E^*_\ell E_\ell (x^0_r) + \sum_s 
i {\pi \over 2} \sum_{\ell < y^1_s} E^*_\ell E_\ell (y^0_s) \eqno(5.4)
$$

and this is the only contribution in the correlation 
functions due to the  field $B$. (The 
appearence of the factor ${1\over 2}$ w.r.t. (5.3) is due to the fact that 
the lattice points $\ell$ giving non--vanishing contributions are at the 
boundary of the support of $B_0$, hence a regularization of the 
$\delta$--functions involved contribute ${1\over 2}$ to the integral.)

\vskip 0.3truecm
\underbar{Remark}
\vskip 0.3truecm

\qquad
Since $B$ has its curvature concentrated at  $\underline{x} =\{x_r\}$ and 
$\underline{y}
= \{y_s\}$, one can view the contribution (5.4) in the action as the effect 
of a disorder field, in the spirit of refs. [23,26,27].

\vskip 0.3truecm

\qquad
Let us sketch the result of the $V_2$--integration, and for more details see,
e.g. [28]. To every field insertion we assign a spin ${1\over 2}$ 
representation of $SU(2)$ with right action  for the creation and 
left action for the annihilation fields, while the Lie--algebra generators 
acting on the $p$--th representation are denoted by $\sigma_a^p, a=x,y,z$. 
Integrating out $V_2$ one obtains a matrix constraint analogous to (5.2):

$$\eqalign{
\epsilon_{\mu\nu} & (\partial^\mu V^\nu + V^\mu V^\nu)_a (z) = \cr
2\pi \chi_{[0, \epsilon]} &
(z^2) \{\sum_s \delta (y^1_s -z^1) 
\delta (y^0_s - z^0) {\sigma_a^s \over 2} \sum_r \delta (x^1_r - z^1) 
\delta (x^0_r - z^0) {\sigma_a^r \over 2} \} . \cr} \eqno(5.5)
$$

A particular matrix valued solution of (5.5) is given by 

$$\eqalign{
& (\bar V_0)_a (z) = 2\pi \chi_{[0,  \epsilon]} (z^2) \{\sum_s \delta(y^0_s - 
z^0) 
\Theta (y^1_s - z^1) {\sigma_a^s \over 2}-\sum_r \delta (x^0_r - z^0) 
\Theta (x^1_r - z^1) {\sigma_a^r \over 2} \}, \cr 
& (\bar V_1)_a (z) =0. \cr}
\eqno(5.6)
$$

The general solution of
(5.5) is obtained from $\bar V_\mu$ by an $SU(2)$ gauge transformation.
If we perform the change of variable (3.17), it is given (at $z^2 = 0$) 
by:

$$\eqalign{
V_0 (z) = & \Bigl( \exp \{- i{\pi \over 2} \sigma_x \sum_{\ell<z^1} 
E^*_\ell E_\ell (z^0)\}
U^\dagger (z) \Bigr) [(\bar V_0)_a (z) \sigma_a + 
\partial_0] \Bigl(U (z)\cdot \cr
& \exp\ \{i{\pi \over 2} \sigma_x \sum_{\ell<z^1} E^*_\ell 
E_\ell (z^0)\} \Bigr), \cr 
V_1 (z) = & \Bigl(\exp\ \{- i {\pi \over 2} \sigma_x \sum_{\ell<z^1} 
E^*_\ell E_\ell (z^0)\} U^\dagger (z) \Bigr) 
\partial_1 \Bigl(U(z) \exp\ \{i {\pi\over 2} \sigma_x \sum_{\ell<z^1} 
E^*_\ell E_\ell (z^0)\} \Bigr). \cr} \eqno(5.7)
$$

\qquad
To understand the effect of the field (5.7),
we use a random walk representation (see [16,20] for details, with an
erratum in [23])
for the correlation functions (5.1) at $\beta \sim + \infty$, 
analogous to the one appearing in (3.20) for the partition function:

$$\eqalign{
\langle & \prod^n_{r=1} E^*_{x^1_r} (x_r^0) \prod^n_{s=1} E_{y^1_s} (y^0_s) 
\prod_r e^{-i \int_{\gamma_{x_r}} B} (P e^{-i 
\int_{\gamma_{x_r}}V})_{1\alpha_r}
\prod_s e^{i \int_{\gamma_{y_s}} B} (P e^{i 
\int_{\gamma_{y_s}} V})_{\beta_s 1} \rangle \cr
= & (\Xi_{t-J})^{-1} \sum_{\pi \in P_n} \prod^n_{k=1} \sum_{\ell_k = 0,1...}
(-1)^{\ell_k} \int_{\scriptstyle \hat\omega_k (x^0_r) = x^1_r \atop 
\scriptstyle \hat\omega_k (y^0_{\pi(s)} + \ell_k \beta) = y^1_{\pi (s)}}
\prod_{k=1}^n d\mu (\hat\omega_k) e^{(\mu+ {J\over 2})(
y^0_{\pi (s)} + \ell_k \beta - x^0_r)} \cdot \cr
\sum_{N=0}^\infty & {e^{\beta(\mu +{J \over 2}) N} \over N!}
\sum_{j_1,..., j_N} \int_{\scriptstyle \omega_p (0)= j_p \atop 
\scriptstyle \omega_p (\beta) = j_p} \prod^N_{p=1} d\mu (\omega_p)
\int {\cal D} U \prod_{\scriptstyle <ij> \in \underline{\omega}_N^\bot}
\Bigl\{\delta (\arg\ (U^\dagger_i U_j)_{[\tilde i][\tilde j]}) \cdot \cr
t & (U^\dagger_i U_j)_{[\tilde i][\tilde j]}\Bigr\} 
\prod_{\scriptstyle \omega \in 
\underline{\omega}_N} \exp\ \Bigl\{\int_{\omega^{\|}} d\tau (U^\dagger 
(\partial_\tau + \bar V_0) U)_{[\tilde \omega (\tau)][\tilde \omega 
(\tau)]} \Bigr\} \ \cdot \cr
\exp\ & \Bigl\{- \sum_{\scriptstyle \omega, \omega^\prime \in 
\underline{\omega}_N} {J \over 2} \int^\beta_0 d\tau \delta_{|\omega(\tau)
- \omega^\prime (\tau)|, 1} \Bigl[|(U^\dagger (\omega (\tau)) 
U(\omega^\prime (\tau))_{[\tilde \omega (\tau)][\tilde 
\omega^\prime(\tau)]} |^2 - {1\over 2} \Bigr] \Bigr\} \cdot \cr
\prod_r & e^{i{\pi \over 2} \tilde x^1_r}
\Bigl( e^{i{\pi \over 2} \tilde x^1_r \sigma_x} U^\dagger_{x^1_r}
(x^0_r) e^{- i {\pi \over 2} \sum_{\ell < x^1_r} \sum_{\scriptstyle \omega \in 
\underline{\omega}_N} (U_\ell \sigma_x U^\dagger_\ell )(x^0_r) 
\delta_{\omega (x^0_r), \ell}} \Bigr)_{1 \alpha_r}\cdot \cr
\prod_s & e^{{\pi\over 2} \tilde y^1_s}
\Bigl( e^{i{\pi \over 2} \sum_{\ell < y^1_s} \sum_{\scriptstyle 
\omega\in \underline{\omega}_N} (U_\ell \sigma_x U^\dagger_\ell) (y^0_s) 
\delta_{\omega(y^0_s),\ell}} U_{y^1_s} (y^0_s) e^{-i {\pi \over 2} \tilde 
y^1_s \sigma_x} \Bigr)_{\beta_s 1},  \cr} \eqno(5.8)
$$

where

$$
\underline{\omega}_N = \{ \hat\omega_1,..., \hat\omega_n, \omega_1,...,
\omega_N \}
$$

and for $j \in {\bf Z}$ 

$$
\tilde j (\tau) = \sum_{\ell<j} \sum_{\scriptstyle \omega \in 
\underline{\omega}_N} \delta_{\omega(\tau),\ell}.
$$

[The contribution of a path--ordered exponential $P e^{i \int_{\gamma_z}
V}$ has been computed by splitting $\gamma_z$ into intervals between 
two consecutive crossing of (the projection in the $0-1$ plane
of) $\gamma_z$ with $\underline{\omega}$: Let $\{z_j \}^{p-1}_{j=1}$
denote the set of the spatial coordinate of the crossing points
ordered from $-\infty$ to $z_1$, and set $z_0 = -\infty, z_p = z$.
Then in (5.8) we find:

$$
P e^{i \int_{\gamma_z} V} = \prod^{p-1}_{j=0} P e^{i \int_{z_j}^{z_{j+1}}
V} = \prod^{p-1}_{j=0} \Bigl(e^{-i {\pi\over 2} j \sigma_x} 
U^\dagger_{z^1_j} (z^0) U_{z^1_{j+1}} (z^0) e^{i {\pi\over 2} (j+1) 
\sigma_x} \Bigr) =
$$
$$
\prod^{p-1}_{j=1} \Bigl(U_{z^1_j} (z^0) e^{i {\pi\over 2}\sigma_x} 
U^\dagger_{z^1_j} (z^0) \Bigr) U_{z^1} (z^0) e^{i {\pi\over 2} \tilde z^1 
\sigma_x}. \quad]
$$

Apart from the presence of $\bar V$ and the exponentials
due to the $\gamma$-strings, the key difference in (5.8)
w.r.t. (3.20) is the appearence of $n$ new random walks
$\hat\omega_k$, starting at times $x^0_r$ at the points
$x^1_r$ and ending 
in the points $y^1_{\pi(s)}$ at time $y^0_{\pi(s)}$;
after wrapping $\ell_k= 0,1...$ times
around the circle of lenght $\beta$ in the time direction;
these paths describe the virtual worldlines of the charged particles
created and annihilated by the insertions of the $E$ and $E^*$
fields. 
Using the techniques of [19,28], one can show that in the presence of $
\bar V$ every crossing of $\underline{\omega}^{\|}_N$ with $\gamma_{x_r}
(\gamma_{y_s})$ contributes to (5.8) a factor  ${\scriptstyle + \atop (-)} i$.

\qquad
Collecting all the factors due to the field $V$ (5.7), we derive that

$$
{\scriptstyle + \atop (-)} i U_j (\tau) \sigma_x 
U_j^\dagger (\tau) \eqno(5.9)
$$

is the contribution of $V$ due to the intersection at site $j$ and time 
$\tau$ of the curves $\gamma_{x_r} (\gamma_{y_s})$ with the virtual 
worldlines $\underline{\omega}_N$ of the charged particles described by 
the field $E$.

\qquad
Combining together (5.4), (5.9) and (3.16) we find that, 
for  the spin singlet 
correlations, the fermion fields can be exactly represented in terms of $U,
E, E^*$, as

$$\eqalign{
\Psi^*_{x^1 \alpha} & (x^0) = E^*_{x^1} e^{i {\pi\over 2} \sum_{\ell<x^1} 
E^*_\ell 
E_\ell (x^0)} \Bigl(U^\dagger_{x^1} (x^0) 
e^{-i{\pi\over 2}  \sum_{\ell<x^1} U_\ell \sigma_x U^\dagger_\ell (x^0) 
E^*_\ell E_\ell 
(x^0)} \Bigr)_{[\tilde x_1]\alpha}, \cr
\Psi_{y^1 \beta} & (y^0) = E_{y^1} (y^0) e^{-i {\pi\over 2} \sum_{\ell <y^1}  
E^*_\ell E_\ell (y^0)}
\Bigl(e^{i {\pi\over 2}  \sum_{\ell<y^1} U_\ell \sigma_x U^\dagger_\ell (y^0) 
E^*_\ell E_\ell (y^0)} U_{y^1} (y^0) \Bigr)_{\beta[\tilde y^1]}. \cr}
\eqno(5.10)
$$

To obtain a more tractable expression to compare with the results obtained
by Bethe--Ansatz and Luttinger liquid techniques,
one applies the MFA  discussed in 
section 4.

\qquad
An important  result of the MFA is  that the spin degrees of freedom 
appear in the 
``squeezed Heisenberg chain", where the Gutzwiller projection can be 
implemented exactly as a single occupancy constraint. The spin fluctuations
$\tilde b_\alpha$ in the squeezed chain can be then converted into fermion 
fields $f_\alpha$ by 

$$
\tilde b_{j\alpha} (\tau) = e^{i \pi \sum_{\ell<j} f^*_{\ell\beta} 
f_{\ell\beta} (\tau)}f_{j\alpha} (\tau). \eqno(5.11)
$$

{}From the Gutzwiller projection we also derive, see (4.14), (4.25):

$$
f^*_{j1} = e^{i\pi j} f_{j2}, \quad f_{j2}^* = e^{i\pi j} f_{j1}. \eqno(5.12)
$$

Denoting quantities in the squeezed chain by $[ \ {\cdot} \ ]^s$, 
we obtain in MFA

$$\eqalign{ 
& \Bigl(e^{i{\pi\over 2} \sum_{\ell<y^1} U_\ell \sigma_x 
U^\dagger_\ell (y^0) 
E^*_\ell E_\ell (y^0)} U_{y^1} (y^0) \Bigr)_{\beta[\tilde y^1]} 
\mathop\sim_{MFA} \cr
& \Bigl[e^{(-)^\beta i{\pi \over 2} \sigma_z \sum_{\ell<y^1} 
b^*_{\ell 1} b^*_{\ell 2} + b_{\ell 1} 
b_{\ell 2}} \tilde b_{y^1 \beta} (y^0) \Bigr]^s \cr
& = \Bigl[e^{(-)^\beta i{\pi\over 2} \sigma_z \sum_{\ell<y^1} (-1)^\ell 
(f_{\ell 1}^* 
f^*_{\ell 2} + f_{\ell 1} f_{\ell 2} )} e^{i\pi \sum_{\ell<y^1} 
f^*_{\ell\alpha} 
f_{\ell\alpha} (y^0)} f_{y^1 \beta}(y^0) \Bigr]^s  \cr
& = \Bigl[e^{-(-)^\beta i {\pi\over 2} \sum_{\ell<y^1} f^*_{\ell\alpha} 
f_{\ell\alpha} 
(y^0)} f_{y^1\beta} (y^0) \Bigr]^s. \cr} \eqno(5.13)
$$

\qquad
To derive the large--scale properties we apply to (5.13) the abelian 
bosonization,
obtaining at long distance, apart from an overall ultraviolet
renormalization (see Appendix A):

$$
\Bigl (e^{i {\pi\over 2} \sum_{\ell<y^1} U_\ell \sigma_x U^\dagger_\ell 
(y^0) E^*_\ell E_\ell (y^0)} 
U_{y^1} (y^0) \Bigr)_{1 \{2 \} \tilde y^1} \mathop\sim_{MFA}
f^{R \{*L \}}_1 (y). \eqno(5.14)
$$

(In (5.14) for brevity we introduced the following
notation: if strings of the form $a \{b\}$ with 
mathematical symbols $a,b$ appearing on both sides of an equation,
the meaning is that the equation is valid either if we use
everywhere the symbols before $\{\cdot \}$, or
if we use everywhere the symbols inside $\{\cdot \}$.)

\qquad
To extract the large scale properties of the charge degrees 
of freedom we apply 
the abelian bosonization also to the field $E$: first  we rewrite it 
(and its conjugate) in terms of left and right movers:

$$
E_j = e^{i\pi (1 - \delta)j} E^L_j + e^{-i\pi(1-\delta)j} E^R_j, 
\eqno(5.15)
$$

then to the corresponding continuum  
fields $E^L(x), E^R(x)$ we apply the 
bosonization scheme rewriting their correlation functions in terms of a 
real scalar field, $\phi_c$, with (euclidean) Thirring--Luttinger action

$$
S(\phi_c) = {1\over 8\pi} \int d^2x [(\partial_0 \phi_c)^2 + v^2_c 
(\partial_1 \phi_c)^2] \eqno(5.16)
$$

and with the identifications

$$
E^R = D_c (x,1): e^{{i\over 2} \phi_c (x)}: {\rm etc}.,
$$

where $v_c \sim t_R$ denotes the charge velocity.

\qquad
Evaluation of the $E-$string in the scaling limit gives

$$
e^{\pm i {\pi\over 2} \sum_{\ell<x^1} E^*_\ell E_\ell (x^0)}  
\sim e^{\pm i {\pi\over
2} \int_{-\infty}^{x^1} : E^{*R} E^R+E^{*L}E^L : (y) dy} 
e^{\pm i {\pi\over 2} (1-\delta) x^1}  
$$
$$\sim e^{\pm {i\over 4} \phi_c (x) 
\pm i {\pi\over 2} (1-\delta) x^1}. \eqno(5.17)
$$

By the same arguments as those used in (4.11--12), one can show 
that the non--local field 
operator reconstructed from  $E(x) e^{\pm i \int^{x^1}_{-\infty}: E^{*R} 
E^R + E^{*L} E^L: (y) dy}$
obeys an abelian braid statistics with statistics parameter $\theta = 
{1\over 4}$.

\qquad
Combining together (5.12--17) we have shown that in MFA the large scale 
behaviour of the spin--singlet correlation functions can be derived using the 
identities:

$$
\Psi^*_{x^1 1 \{2 \}} (x^0) \mathop\sim_{MFA} \Bigl(e^{-1 \{-3\}
i{\pi \over 2} (1-\delta) x^1}
E^{*L} (x) + e^{3 \{1\} i {\pi\over 2} (1-\delta) x^1} E^{*R} (x) \Bigr)
\cdot
$$
$$
e^{+ \{-\} i {\pi\over 2} \int^{x^1}_{-\infty} : E^{*R} E^R + 
E^{*L} E^L : (z) dz} f_1^{*R \{L\}} (x) \sim
$$
$$
D_c (x, -1) (e^{1 \{3 \} i {\pi\over 2} (1-\delta) x^1} : 
e^{3 \{1\} {i \over 4} \phi_c (x)}: + e^{3 \{1\} 
i {\pi \over 2}  (1-\delta) x^1}
: e^{-3  \{-1\}  {i \over 4} \phi_c (x)} :) \cdot
$$
$$
D_- (x, - \{+\} {1\over 2}) : e^{{i\over 2} \phi_- (x)}:
$$
$$
\Psi_{y^1 1 \{2\}} (y^0) \mathop\sim_{MFA} \Bigl(e^{1 \{3\} i 
{\pi\over 2} (1-\delta) y^1} E^L (y) + 
e^{-3  \{-1\} i{\pi \over 2} (1-\delta) y^1} E^R (y) \Bigr) \cdot
$$
$$
e^{-\{+\} i {\pi\over 2} \int^{y^1}_{-\infty} : E^{*R} E^R + E^{*L} 
E^L : (z) dz} f_1^{R \{*L\}} (y) \sim
$$
$$
D_c (y,+1) (e^{1 \{3\} i {\pi\over 2} (1-\delta) y^1}: e^{-1 
\{-3\} {i\over 4} \phi_c (y)} :
+ e^{-3 \{-1\} i{\pi \over 2} (1-\delta) y^1}: e^{3 \{1\} 
{i\over 4} \phi_c (y)}:) \cdot
$$
$$
D_- (y, + \{-\}  {1\over 2} )
: e^{- {i\over 2} \phi_- (x)}: \eqno(5.18)
$$

According to the general results of $U(1) \times SU(2)$ Chern--Simons 
bosonization, (see (3.13)), the original fermion field can be 
decomposed into two 
non--local semion fields,  $U(1) \times SU(2)$--gauge invariant:
$E(\gamma_x|B), \Sigma_\alpha (\gamma_x|V)$
which one may call holon and spinon field, respectively. Equation (5.18) 
proves that for large scales, in the MFA one can identify $E(\gamma_x|B)$ as 
the 1D--semion field $E(x) e^{i \int^{x1}_{-\infty}: E^{*R} E^R + E^{*L} 
E^L:}$ and $\Sigma_{1 \{2\}} (\gamma_x|V)$ as the 1D 
chiral semion field 
$f_1^{R \{*L\}}$, a Gutzwiller projected chiral fermion field 
of the squeezed Heisenberg chain.

\qquad
In  Appendix B  
we show that the formulas (5.18) when applied to the correlation 
functions of the 1D $t-J$ model in the regime $t >> T$, indeed reproduce 
correctly large scale behaviours identical to those obtained with 
Bethe--Ansatz and Luttinger liquid techniques extrapolated from the large 
$U$ Hubbard model [5] and the $t=J$, $t-J$ model [6]. 

\qquad
To conclude, we have shown that one can obtain the correct large 
scale behavior of the 1D $t-J$ model by simply using 
a mean field 
theory treatment of the dimensional reduction of the $U(1) \times SU(2)$ 
Chern--Simons bosonization.  
Moreover, we have shown that the $U(1) \times SU(2)$ Chern--Simons 
bosonization is  the most natural 
mathematical framework for  describing the spin--charge decomposition of the 
electron 
field in terms of semionic fields. This shows that the key physical
properties of the 1D $t-J$ model are captured not by an arbitrary 
spin--charge separation scheme 
(in fact slave fermion and slave boson approaches
failed to reproduce the correct large scale behaviour 
of the correlation functions), but rather by a specific semionic form of the 
spin--charge separation. This gives rise to our hope that the power of this 
formalism and the underlying physical intuition will survive in 2D.

\vskip 0.5truecm
{\bf Appendix A}
\vskip 0.3truecm

\qquad
We first fermionize the system (4.8). Since the spin components of 
$\tilde b_\alpha$ are coupled via the Gutzwiller constraint, one cannot apply 
to them independent Jordan--Wigner transformations. To derive the correct
transformation we use once more the reduction from a 2D system. Coupling 
the $\tilde b$ field to a $U(1)$ gauge field $B$ with action $S_{c.s.} (B)$
in 2D is known to convert $\tilde b$ to a fermion field $f$. Integrating 
out $B_0$ we get the constraint

$$
\epsilon_{\mu\nu} \partial^\mu B^\nu (x) =2\pi \sum_j \tilde b_{j\alpha}^* 
\tilde b_{j\alpha} (x^0) \delta (x^1 -j), \quad  \mu,\nu=1,2.
$$

In the Landau gauge, $\partial^\mu B_\mu =0$, it can be solved by

$$
B_\mu (x) = \sum_j \tilde b_{j\alpha}^* \tilde b_{j\alpha} 
(x^0) \partial_\mu \arg\ (x^1 -j).
$$

Reducing this formula to 1D we find

$$
B_0 (x) = 0, \qquad \qquad
B_1 (x) = \pi \sum_j \tilde b_{j\alpha}^* \tilde b_{j\alpha}  (x^0) \partial_1
\Theta (x^1-j).
$$

Applying the Gutzwiller constraint we obtain 

$$
B_1 (x) = \pi \sum_j \delta (x^1 - j), 
$$

so that $e^{i\int_{<ij>}B } = e^{i \pi}$ and the action for fermions 
reads as

$$
S(f,f^*) = \int d\tau \sum_i f^*_i \partial_\tau f_i - \sum_{<ij>} \tilde J_R 
(f^*_i f_j + h.c.) \eqno(A.1)
$$

with the constraint

$$
f_{j\alpha}^* f_{j\alpha} =1, \eqno(A.2)
$$

i.e. it describes exactly a system of spin ${1\over 2}$ non-relativistic free 
fermions, Gutzwiller projected. 

\qquad
It is well known [29] that the large--scale properties of the spin ${1\over
2}$ quantum Heisenberg chain can be described in terms of a real scalar 
field $\varphi$ with Luttinger--Thirring action 

$$
S_\xi (\varphi)= {\xi\over 8\pi} \int [(\partial_0 \varphi)^2 + 
v^2_s (\partial_1 \varphi)^2], \eqno(A.3)
$$

where $\xi =2$ and $v_s$ denotes the spin velocity.
(Here we use the convention in which a mass term would be described by $:
\cos\ \varphi:$.)

\qquad
Let us now  show that we indeed recover this result starting from the mean field
model (A.1).
To analyze the large scale properties, following [30], we first introduce 
the decomposition of $f$ in right and left movers in a lattice labelled by 
sites in $(2 {\bf Z} + {1\over 2})$:

$$\eqalign{
f_{2n\alpha} = & i^{2n} f_{(2n +{1\over 2}) \alpha}^L + (-i)^{2n} 
f^R_{(2n+{1\over 2})\alpha}, \cr
f_{2n+1\alpha} = & i^{2n+1} f^L_{(2n+{1\over 2})\alpha} + (-i)^{2n+1}
f^R_{(2n+ {1\over 2})\alpha}. \cr} \eqno(A.4)
$$

The Gutzwiller constraints are   given by:

$$\eqalign{
f^{* L}_{2n + {1\over 2} \alpha} & f^L_{2n+{1\over 2} \alpha} + f^{* 
R}_{2n +{1\over 2} \alpha} f^R_{2n+{1\over 2} \alpha} = 1, \cr
f^{* R}_{2n+{1\over 2}\alpha} & f^L_{2n+{1\over 2} \alpha}+ f^{* 
L}_{2n +{1\over 2} \alpha} f^R_{2n+{1\over 2}\alpha} =0. \cr} \eqno(A.5)
$$

The fields $f^L, f^R$ are then assumed to have a good continuum limit with 
linearized dispersion relations, resulting in a large--scale continuum 
action given, before the constraint is implemented, by

$$
S_{M F} (f, f^*)= \int d^2 x \Big[ f^{* 
R}_\alpha(\partial_0 + i v_s \partial_1) f^R_\alpha + 
f_\alpha^{* L} (\partial_0 - i v_s \partial_1) f^L_\alpha\Big].\eqno(A.6)
$$

\qquad
Introducing two real scalar field $\phi_\alpha, 
\alpha =1,2$ and applying the standard 
abelian 1D bosonization (proved in [23,31] to be a special version
of the 
duality transformation), we obtain a Luttinger--Thirring action

$$
S_{M F} (\phi_\alpha) = \sum_{\alpha=1}^2 S_1 (\phi_\alpha), \eqno(A.7)
$$

and the bosonization formulas for fields [23,27]:

$$\eqalign{
f^R_{\alpha} (x) \sim & (2\pi)^{-{1\over 4}} D_\alpha (x,1): 
e^{-{i\over 2} \phi_\alpha(x)}:, \cr
f^{* R}_{\alpha}(x) \sim & (2\pi)^{-{1\over 4}} D_\alpha (x,-1) : e^{{i\over 2} 
\phi_\alpha (x)}: ,\cr
f^L_\alpha (x) \sim & (2\pi)^{-{1\over 4}} D_\alpha (x,1): 
e^{{i\over 2} \phi_\alpha (x)} :, \cr
f^{* L}_\alpha (x) \sim & (2\pi)^{-{1\over 4}} D_\alpha (x, -1): 
e^{- {i\over 2} \phi_\alpha (x)}:, \cr} \eqno(A.8)
$$

where $D_\alpha (x, \pm 1)$ is a disorder field and we adopted the 
notations
of reference [23], to which we refer for more details.
[In (A.8) $:e^{i\zeta \phi (x)} :, \zeta \in {\bf R}$ denotes
the normal ordered exponential, defined as follows: let $\delta^\kappa_x$ 
be a regularization of the Dirac $\delta$--function $\delta_x$, with u.v.
regulator $\kappa$, i.e. $\delta^\kappa_x {\rightarrow \atop \kappa 
\uparrow \infty} \delta_x$; then

$$
: e^{i \zeta \phi(x)}: \equiv  \lim_{\kappa \uparrow \infty} e^{i\zeta \phi 
(\delta^\kappa_x)} (2\pi)^{\zeta^2} e^{-{2\pi \zeta^2 \over \xi} 
(\delta^\kappa_x, \Delta^{-1} \delta^\kappa_x)} \eqno(A.9)
$$

with 

$$
\Delta^{-1} (x,y) \equiv {1\over 2\pi} \ln \sqrt{(x^0 - y^0)^2 + v^2_s
(x^1 - y^1)^2}.
$$

Formally, we rewrite (A.9) as

$$
: e^{i\zeta \phi (x)}: = e^{i\zeta \phi(x)} (2\pi)^{\zeta^2}
e^{-{2\pi \zeta^2 \over \xi} 
\Delta^{-1} (x,x)}. \qquad] \eqno(A.10)
$$

In the model with action (A.3) the expectation values of products of
disorder fields and exponentials are given by

$$
< \prod_j D(x_j, \zeta_j) e^{i\int \phi(x) f(y) d y} >
$$
$$
= \cases{0, \ {\rm if} \ \sum_j \zeta_j \not = 0 \ {\rm or} \ \hat f (0) \not =
0 \quad {\rm for} \ f \ {\rm real}, \cr
e^{{\xi \over 2} \sum_{i <j} \zeta_i \zeta_j \ln |x_i - x_j|} 
e^{\int d^2 x d^2 y f(x) \ln |x - y| f (y)} e^{i\sum_j \int d^2 x
f(x) \arg\ [(x^0 - x^0_j)+i v_s(x^1 - x^1_j)]}, \cr} \eqno(A.11)
$$

where $|z - w| \equiv \sqrt{(z^0 - w^0)^2 + v_s^2 (z^1 - w^1)^2}$.

One can implement the Gutzwiller constraint exactly, defining

$$
\phi_\pm = {\phi_1 \pm \phi_2 \over 2}.
$$

Then, with the notations of (A.3 ), we have

$$
S_{MF} (\phi_+, \phi_-) = S_2 (\phi_+) +S_2 (\phi_-) \eqno(A.12)
$$

and, e.g.,

$$
f^R_1 (x) = D_+ (x, {1\over 2}) D_-(x, {1\over 2}): e^{-{i\over 2} \phi_+ 
(x)} : e^{-{i\over 2} \phi_- (x)}: 
$$

etc, so that (taking care of the normal ordering, see [27]) the 
constraints (A.5) become

$$\eqalign{
{1 \over 2\pi}: & \partial_1 \phi_+ :=0, \cr
:\cos\ & (\phi_+ + {\pi\over 2}): :\cos\ \phi_- :=0, \cr}
\eqno(A.13)
$$

solved by 

$$
\phi_+ = {\rm const}, \quad  :e^{i \phi_+} : =1. \eqno(A.14)
$$

As a conseguence $S_{MF} (\phi) = S_2 (\phi_-)$, i.e. we recover
eq. (4.9) and we obtain the bosonization formulas (4.10).

\qquad
{}From equations (4.10) and (A.11) we derive, e.g.,

$$\eqalign{
& \langle f^*_\uparrow (x) f_\uparrow (y) \rangle = 
e^{i{\pi \over 2} (x^1 - y^1)} \langle f^{*R}_1 (x) f_1^R (y) 
\rangle + 
e^{-i{\pi\over 2} (x^1 -y^1)} \langle f^{*L}_1 (x) f_1^L (y) 
\rangle = \cr 
& {e^{i {\pi \over 2} (x^1 - y^1)} \over \sqrt{(x^1 -y^1) - i v_s(x^0 - 
y^0)}} + {e^{-i{\pi\over 2} (x^1 - y^1)} \over \sqrt{(x^1 -y^1)+ i v_s (x^0 
- y^0)}}. \cr}
$$

Therefore the mean field treatment  of Gutzwiller projected free electrons 
reproduces 
exactly the large scale properties of the Heisenberg chain.

\underbar{Remark}:
Suppose we keep the hard--core constraint for the individual fields 
$\tilde b_\alpha$, but perform  a mean field treatment of the remaining 
Gutzwiller constraint, then, since the two components of $\tilde b_\alpha$ 
are not any more coupled we can fermionize them by separate Jordan--Wigner 
transformations. To the corresponding fermionic field $f_\alpha$
one can apply the same treatment as before, but since the Gutzwiller
constraint disappeared, our bosonization formulas are simply eqs.(4.18). This
procedure reproduces the approximate results of [7]. Therefore, the treatment
of the constraint should be exact, and only afterwards one can use  MFA.

\qquad
Finally, let us apply equations (4.10) and (A.9) to prove
(5.14): we rewrite (5.13) as

$$\eqalign{
& e^{-(-)^\beta i {\pi\over 2} y(1-\delta)} e^{{i\over 2} \phi_+(y)} D_+ (y,
{1\over 2}) D_- (y, {1\over 2}) \cdot \cr
& \Bigl(e^{-i {\pi\over 2} (1-\delta) y^1}: e^{-{i\over 2} 
\phi_+ (y)}: :e^{(-)^\beta {i\over 2} \phi_- (y)}: + e^{+i{\pi\over 2} 
(1-\delta) y^1} 
e^{{i\over 2} \phi_+ (y)}: : e^{-(-)^\beta {i\over 2} \phi_- (y)}: \Bigr) 
\cr} \eqno(A.15)
$$

with the Gutzwiller projection implemented imposing equation (A.14).
Formally, see (A.9--A.10), we have

$$\eqalign{
e^{{i\over 2} \phi_+ (y)}: & e^{-{i\over 2} \phi_+ (y)}: \sim e^{-{1\over 8} 
\Delta^{-1} (y,y)}, \cr
e^{{i\over 2} \phi_+ (y)}: & e^{{i\over 2} \phi_+ (y)}: \sim e^{{3\over 8} 
\Delta^{-1} (y,y)} : e^{i\phi_+ (y)}: = e^{{3\over 8} \Delta^{-1} (y,y)} .
\cr}
$$

\qquad
Therefore, the ratio of the coefficients of the second to the first 
term in (A.15) tends to $0$ as the U.V. cutoff (assumed to be the inverse 
of the scale parameter) is removed.
Recalling eq.(4.10), at large scale we recover (5.14).

\vskip 0.5truecm
{\bf Appendix B}
\vskip 0.3truecm

Using (5.18) and Appendix A we compute the large scale behaviour of

1) density--density correlation function:

$$
\langle : \Psi^*_\alpha \Psi_\alpha: (x) : \Psi^*_\beta
\Psi_\beta : (y) \rangle \mathop\sim_{MFA} 
\langle: E^* E: (x) : E^* E: (y) \rangle 
$$
$$
\sim \langle {1\over 2\pi} :\partial_1 \phi_c :(x) {1\over 2\pi} :
\partial_1 \phi_c : (y) \rangle 
$$
$$
+ \langle : e^{i\phi_c (x)}: : e^{-i\phi_c (y)} \rangle
e^{-i 2\pi (1-\delta) (x^1-y^1)} + \langle: e^{-i \phi_c (x)}: : e^{i\phi_c 
(y)}: \rangle e^{i 2\pi (1-\delta) (x^1-y^1)}
$$
$$
= -{1\over 4\pi^2} \Bigl[{1\over ((x^1 - y^1) -iv_c (x^0-y^0))^2} + {1\over
((x^1 - y^1)+ iv_c(x^0-y^0))^2} \Bigr] 
$$
$$
+ {1\over 2\pi^2}  {\cos [2\pi 
(1-\delta)(x^1-y^1)] \over (x^1-y^1)^2+ v^2_c (x^0-y^0)^2}; \eqno(B.1)
$$

2) spin--spin correlation function

$$
\langle \Psi^*_\alpha {\vec \sigma_{\alpha\beta} \over 2} \Psi_\beta (x) 
\Psi^*_\gamma {\vec \sigma_{\gamma\delta} \over 2} \Psi_\delta (y) 
\rangle =
$$
$$
= {1\over 2} \langle \Psi^*_\alpha (x) \Psi_\beta (x) \Psi^0_\beta(y) 
\Psi_\alpha (x) \rangle
-{1\over 4} \langle \Psi^*_\alpha (x) \Psi_\alpha(x) \Psi^0_\beta (y) 
\Psi_\beta (y) \rangle \mathop\sim_{MFA} 
$$
$$
{1\over 2} \langle E^* E(x) f^{*R}_1 f^R_1 (x) E^* E(y) f^*_1 
f^R_1 (y) \rangle
+ {1\over 2} \langle E^* E(x) f^{*L}_1 f^L_1 (x) E^* E(y) f^{*L}_1 f^L_1 (y) 
\rangle
$$
$$
+ {1\over 2} \langle E^* E(x) e^{- i\pi \int^{x^1}_{-\infty} : E^{*R} E^R 
+E^{*L} E^L : (z) dz} f^R_1 f^L_1 (x) E^* E(y) e^{i\pi
\int^{y^1}_{-\infty}: E^{*R} E^R + E^{*L} E^L: (z) dz} f^{*R}_1 f^{*R}_1 (y) 
\rangle
$$
$$
+{1\over 2} \langle E^* E(x) e^{i\pi \int^{x^1}_{-\infty}: E^{*R} E^R +
E^{*L} E^L: (z) dz} f^{*R}_1 f^{*L}_1 (x) E^* E(y) e^{-i\pi 
\int^{y^1}_{-\infty} : E^{*R} E^R + E^{*L} E^L : (z) dz} f^R_1 f^L_1 (y) 
\rangle 
$$
$$
-{1\over 4} \langle E^* E(x) E^* E(y) \rangle
\sim {\delta^2 \over 2} \langle e^{i\pi \int^{y^1}_{x^1}: E^{*R} E^R + E^{*L}
E^L: (z) dz} \rangle e^{i\pi (1-\delta) (y^1-x^1)} \cdot
$$
$$
\langle f^{*R}_1 (x) f^R_1 (y) \rangle \langle f^{*L}_1 (x) f^L_1(y) 
\rangle + \langle e^{-i\pi \int^{y^1}_{x^1}: E^{*R} E^R + E^{*L} E^L :(z) dz}
\rangle \cdot
$$
$$
e^{-i\pi(1-\delta) (y^1-x^1)} \langle f^R (x) f^{*R} (y) \rangle \langle f^L
(x) f^{xL} (y) \rangle
$$
$$
= \delta^2 {\cos\ [(\pi (1-\delta) (y^1-x^1))] \over ((x^1 -y^1)^2 + 
v^2_s 
(x^0-y^0)^2)^{1\over 2} ((x^1 -y^1)^2 + v^2_c (x^0 -y^0)^2)^{1\over 4}},
\eqno(B.2)
$$

where in the third equality we use $E^* E (x) = : E^* E: (x) + \delta$
and we neglect the first term as subleading;

\vskip 0.3truecm
3) electron--electron correlation function

$$
\langle \Psi_\mu^* (x) \Psi_\nu (y) \rangle = {\delta_{\mu\nu} \over
2} \langle \Psi^*_\alpha (x) \Psi_\alpha (y) \rangle; 
$$
$$
\langle \Psi^*_\alpha (x) \Psi_\alpha (y) \rangle
\mathop\sim_{MFA}  \langle (E^{*L}(x) E^L (y) e^{-i{\pi \over 2} (1-\delta) (x^1-y^1)} + 
E^{*R} (x) E^R (y)
e^{3i {\pi \over 2} (1-\delta) (x^1-y^1)}) \cdot 
$$
$$
e^{-i{\pi \over 2} \int^{y^1}_{x^1}: 
E^{*R} E^R + E^{*L} E^L : (z) dz} f^{*R}_1 (x) f^R_1 (y) \rangle
$$
$$
+ \langle (E^{*L} (x) E^L (y) e^{-{3\over 2} i\pi (1-\delta) (x^1-y^1)}+ E^{*R}
(x) E^R (y) e^{i {\pi\over 2}(1-\delta) (x^1-y^1)})\cdot
$$
$$
e^{i{\pi\over 2} \int_{x^1}^{y^1} : E^{*R} E^R + E^{*L} E^L : (z) dz}
f_1^L (x) f^{*L}_1 (y) \rangle = \langle D_c (x, -1) D_c (y, 1)
D_- (x, -{1\over 2}) D_- (y, {1\over 2}) 
$$
$$
\Bigl[e^{-i{\pi \over 2} (1-\delta)(x^1 - y^1)}: e^{{i \over 4} \phi_c 
(x)}: : e^{-{i \over 4} \phi_c (y)}: + e^{i 3{\pi \over 2} (1-\delta)
(x^1 - y^1)}: e^{-i{3 \over 4} \phi_c (x)}: : e^{i{3\over 4} \phi_c (y)}:
\Bigr] \rangle + h.c.
$$
$$
= {1\over [(x^1-y^1) + i v_s (x^0 - y^0)]^{1\over 2}} 
{1\over [(x^1
- y^1)^2 + v_c^2 (x^0 - y^0)^2]^{1\over 16}}\cdot 
$$
$$
\Bigl[{e^{i{\pi \over 2}
(1-\delta) (x^1 - y^1)} \over [(x^1 - y^1) + i v_c (x^0 - y^0)]^{1\over 2}}
+ {e^{i3 {\pi \over 2}( 1-\delta) (x^1-y^1)} \over [(x^1 - y^1) + i v_c
(x^0 - y^0)]^{3\over 2}} \Bigr]+ h.c. \eqno(B.3)
$$

We see that  the large scale behaviour of the 
correlation functions is of the form (1.2) and a comparison
of the values of  $n, \alpha^\pm_c$, $\alpha^\pm_s$ with those 
found in [5] and [6] shows that they agree exactly with the result
obtained for the large $U$ Hubbard model and the $t-J$ model at $t=J$, 
extrapolated to the region $t>>J$.

\vskip 0.5truecm

{\bf Acknowledgments.} Useful discussions with J. Fr\"ohlich are 
gratefully acknowledged.

\vskip 0.5truecm

\vfill\eject

{\bf References}

\vskip 0.3truecm

\line{[1] T.M. Rice, F. Mila, F.C. Zhang, Phil. Trans. R. Soc. 
London  {\bf A334} (1991) \hfil}
\line{ \hskip .8truecm  459 and references  therein.\hfil}

\smallskip

[2] E.H. Lieb, F.Y. Wu, Phys. Rev. Lett. {\bf 20} (1968) 1445.

\smallskip

[3] M. Ogata, H. Shiba, Phys. Rev. {\bf B41} (1990) 2326.

\smallskip

[4] J. Carmelo, D. Baeriswyl, Phys. Rev. {\bf B37} (1988) 7541

\smallskip

\line{[5] H. J. Schulz, Phys. Rev. Lett. {\bf 64} (1990) 283; \hfil}
\line{\hskip .8truecm Y. Ren, P.W. Anderson, Phys. Rev. {\bf B48} (1993) 16662.
\hfil}

\smallskip

\line{[6] H. Frahm, V.E. Korepin Phys. Rev. {\bf B42} (1990) 10553;\hfil}
\line{\hskip .8truecm   N. Kawakami, S.K. Yang, J. Phys. {\bf C3} (1991) 5983.\hfil}

\smallskip

\line{[7] Z.Y. Weng, D.N. Sheng, C.S. Ting and Z.B. Su, Phys. Rev. Lett. {\bf 
67} (1991)\hfil}
\line{\hskip .8truecm  3318; Phys. Rev. {\bf B45} (1992) 7850.\hfil}

\smallskip

\line{[8] P.W. Anderson, Science {\bf 235} (1987) 1196; in ``Frontiers and 
Borderlines\hfil}
\line{\hskip .8truecm in Many--Particle Physics", edited by R.A. Broglia et al. 
(North--Holland,\hfil}
\line{\hskip .8truecm  Amsterdam 1988).\hfil}

\smallskip

\line{[9] A.P. Arovas, A. Auerbach, Phys. Rev. {\bf B38} (1988) 316; \hfil}
\line{\hskip .8truecm  D. Yoshioka, J. Phys. Soc. Jp. {\bf 58} (1989)  1516 and references 
therein.\hfil}

\smallskip

[10] L.B. Ioffe, A.I. Larkin, Phys. Rev. {\bf B39} (1989)  
8988 and references therein.

\smallskip

[11] R.B. Laughlin, Science {\bf 242} (1988) 525; Int. J. Mod. Phys. {\bf 
B5} (1991) 1587.

\smallskip

[12] A. Lerda, S. Sciuto, Nucl. Phys. {\bf B410} (1993) 577.

\smallskip

\line{[13] ``Fractional Statistics and Anyon Superconductivity", edited 
by \hfil}
\line{\hskip .92truecm F. Wilczek (World Scientific, Singapore 1990).\hfil}

\smallskip

\line{[14] G.W. Semenoff, Phys. Rev. Lett. {\bf 61} (1988) 517;\hfil}
\line{ \hskip .8truecm  E. Fradkin, 
Phys. Rev. Lett. {\bf 63} (1989) 322;\hfil}
\line{ \hskip .8truecm  M. L\"uscher, Nucl. Phys. {\bf B326}
(1989) 557.\hfil}

\smallskip

[15] J. Fr\"ohlich, T. Kerler, P.A. Marchetti, Nucl. Phys. {\bf 374} 
(1992) 511.

\smallskip

[16] J. Fr\"ohlich, P.A. Marchetti, Phys. Rev. {\bf B46} (1992) 6535.

\smallskip

[17] P.A. Marchetti, Nucl. Phys. B (Proc. Suppl.) {\bf 33C} (1993) 134.

\smallskip

[18] B. Schroer, J.A. Swieca, Nucl. Phys. {\bf B121} (1977) 505.

\smallskip

\line{[19]  E. Witten, Commun. Math. Phys. {\bf 121} (1989) 351;\hfil}
\line{ \hskip .8truecm  J. Fr\"ohlich, C. King, Int. J. Mod. Phys. {\bf A4}
 (1989) 5321.\hfil}

\smallskip

\line{[20] J. Ginibre, Commun. Math. Phys. {\bf 10} (1968) 140;\hfil}
\line{ \hskip .8truecm  J. Ginibre, C. Gruber, ibid. {\bf 11} (1068) 198.\hfil}

\smallskip

[21] I. Affleck, J.B. Marston, Phys. Rev. {\bf B37} (1988) 3774.

\smallskip

\line{[22]  J.W. Negele, H. Orland ``Quantum Many Particle Systems"\hfil}
\line{ \hskip .8truecm  (Addison--Wesley, New York 1988).\hfil}

\smallskip

[23] J. Fr\"ohlich, R. G\"otschmann, P.A. Marchetti, J. Phys. {\bf A28} 
(1995) 1169.

\smallskip

\line{[24]  J. Glimm, A. Jaffe ``Quantum Physics. A Functional Integral 
Point of View" \hfil}
\line{ \hskip .8truecm (Springer--Verlag, New York, 1981).\hfil}

\smallskip

\line{[25] A.B. Zamolodchikov, V. Fateev, Sov. Phys. JETP {\bf 62} (1985) 251;
\hfil}
\line{ \hskip .8truecm  {\bf 63} (1985) 913.\hfil}

\smallskip

\line{[26] E.C. Marino, B. Schroer, J.A. Swieca, Nucl. Phys. {\bf B200} [FS4]
(1982) 473;\hfil}
\line{ \hskip .8truecm  J. Fr\"ohlich, P.A. Marchetti, Commun Math. Phys.  {\bf 112} (1987) 343;
{\bf 121} \hfil}
\line{ \hskip .8truecm (1989) 177.\hfil}

\smallskip

[27] J. Fr\"ohlich, P.A. Marchetti, Commun. Math. Phys. {\bf 116} (1987) 127.

\smallskip

\line{[28] E. Guadagnini, M. Martellini and M. Mintchev, Nucl. Phys. {\bf B330}
\hfil}
\line{ \hskip .8truecm (1990) 575.\hfil}

\smallskip

\line{[29] See, e.g., E. Fradkin ``Field theories of condensed matter systems"
\hfil}
\line{ \hskip .8truecm 
(Addison--Wesley, New York 1991).\hfil}

\smallskip

[30] I. Affleck, Nucl. Phys. {\bf B265} (1986) 409.

\smallskip

\line{[31] G. Burgess, F. Quevedo, Nucl. Phys. {\bf B421} (1994) 373;\hfil}
\line{ \hskip .8truecm P.A. 
Marchetti ``Bosonization and duality in condensed matter systems",\hfil}
\line{\hskip .8truecm to appear
in Proceedings of ``Common Trends in Condensed Matter  and \hfil}
\line{\hskip .8truecm  High Energy 
Physics" Chia 1995 --Preprint  hep-th 9511100.\hfil}

\bye